\documentclass[11pt]{article}          
\usepackage[english]{babel}
\usepackage{latexsym}
\makeatletter
\def\@begintheorem#1#2{\trivlist%
 \item[\hskip \labelsep{\sffamily\bfseries #2\ #1}]\itshape}
\newtheorem{teo}{Theorem}[section]
\newtheorem{defi}[teo]{Definition}
\newtheorem{cor}[teo]{Corollary}
\newtheorem{lem}[teo]{Lemma}
\newtheorem{pro}[teo]{Proposition}
\newtheorem{_rem}[teo]{Remark}
\newtheorem{_eje}[teo]{Example}

\newenvironment{rem}{\def\@begintheorem##1##2{\trivlist%
 \item[\hskip\labelsep{\sffamily\bfseries ##2\ ##1}]}\begin{_rem}}{\end{_rem}}

\makeatother

\newenvironment{beweis}{{\em Proof:}}{\hfill $\rule{2mm}{2mm}$
\vspace{3mm}

}

\DeclareMathAlphabet{\Ma}{U}{msa}{m}{n}
\DeclareMathAlphabet{\Mb}{U}{msb}{m}{n}
\DeclareMathAlphabet{\Meuf}{U}{euf}{m}{n}

\def\got#1{\Meuf{#1}}

\DeclareSymbolFont{ASMa}{U}{msa}{m}{n}
\DeclareSymbolFont{ASMb}{U}{msb}{m}{n}
\DeclareMathSymbol{\hrist}{\mathord}{ASMa}{"16}
\DeclareMathSymbol{\varkappa}{\mathalpha}{ASMb}{"7B}
\DeclareMathSymbol{\CrPr}{\mathord}{ASMb}{"6F}

\newfont{\EinsFont}{cmr7 scaled 1070}
\def\EINS{{\mathchoice{
 \mbox{\unitlength1cm\begin{picture}(.25,.2)\put(0,0){$1$}%
 \put(0.105,0){{\mbox{\fontfamily{cmr}\upshape\small l}}}\end{picture}}}{%
 \mbox{\unitlength1cm\begin{picture}(.25,.2)\put(0,0){$1$}%
 \put(0.105,0){{\mbox{\fontfamily{cmr}\upshape\small l}}}\end{picture}}}{%
 \mbox{\unitlength1cm\begin{picture}(.18,.15)\put(0,0){$\scriptstyle 1$}%
 \put(0.07,0){{\mbox{\fontfamily{cmr}\upshape\EinsFont l}}}\end{picture}}}{%
 \mbox{\unitlength1cm\begin{picture}(.18,.15)\put(0,0){$\scriptstyle 1$}%
 \put(0.07,0){{\mbox{\fontfamily{cmr}\upshape\EinsFont l}}}\end{picture}}}}}

\def\restriction{{\mathchoice{
 \mbox{\unitlength1cm\begin{picture}(.2,.4)%
  \bezier{5}(.07,.3)(.1,.27)(.13,.24)%
  \put(.07,.35){\line(0,-1){.5}}\end{picture}}}{
 \mbox{\unitlength1cm\begin{picture}(.2,.4)%
  \bezier{5}(.07,.3)(.1,.27)(.13,.24)%
  \put(.07,.35){\line(0,-1){.5}}\end{picture}}}{
  \hrist}{\hrist}}}

  \def\al #1.{{\mathcal{#1}}}
  \def\ot #1.{{\got{#1}}}

  \def\ccr #1,#2.{\overline{\Delta(#1,\,#2)}}

  \def\b #1.{{\bf #1}}
  \def\cross#1.{\mathrel{\mathop{\times}\limits_{#1}}}
  \def\C{\Mb{C}}
  \def\N{\Mb{N}}
  
  \def\R{\Mb{R}}
  \def\Z{\Mb{Z}}
  \def\T{\Mb{T}}

  \def\wh{\widehat}
  \def\wt{\widetilde}

  \def\cross #1.{\mathrel{\raise 3pt\hbox{$\mathop\times\limits_{#1}$}}}
\def\set #1,#2.{\left\{\,#1\;\bigm|\;#2\,\right\}}
\def\b #1.{{\bf #1}}
\def\ker{{\rm Ker}\,}
\def\aut{{\rm Aut}\,}
\def\out{{\rm Out}\,}
\def\ob{{\rm Ob}\,}
\def\endo{{\rm End}\,}
\def\spec{{\rm spec}\;}

\def\Ad{{\rm Ad}\,}
\def\ad{{\rm ad}\,}
\def\tr{{\rm Tr}\,}
\def\autx{{\rm Aut}_\xi}
\def\ol #1.{\overline{#1}}
\def\rn#1.{\romannumeral{#1}}
\def\chop{\hfill\break}
\def\rest{\restriction}
\def\un{\EINS}
\def\spa{{\rm span}}
\def\csp{\hbox{\rm clo-span}}
\def\clo{{\rm clo}}
\def\s #1.{_{\smash{\lower2pt\hbox{\mathsurround=0pt $\scriptstyle #1$}}\mathsurround=3pt}}
\def\bra #1,#2.{{\left\langle #1,\,#2\right\rangle_{\al A.}}}

\def\XP#1!{\renewcommand{\baselinestretch}{.7}\marginpar{{\footnotesize #1}\hfil}
\renewcommand{\baselinestretch}{1.5}}
\def\XB{\marginpar{
{\footnotesize\bf Change~starts-----}\lower 11pt\hbox{\mathsurround=0pt$
\!\!\displaystyle{
\Bigg\downarrow}$\mathsurround=3pt}}}
\def\XE{\marginpar{{\footnotesize\bf Change~ends-----}\raise 10pt\hbox{\mathsurround=0pt$ 
\!\!\displaystyle{
\Bigg\downarrow}$\mathsurround=3pt}}}
\def\HS{{\{\al F.,\,\al G.\}}}

\title{\bf Superselection in the presence of constraints.}

\author{
  {\sc Hellmut Baumgaertel}   \\
 {\footnotesize
    Mathematical Institute, University of Potsdam,}    \\ 
 {\footnotesize
   Am Neuen Palais 10, Postfach 601~553,}   \\
 {\footnotesize
   D--14415 Potsdam, Germany.}    \\
 {\footnotesize
   E-mail: baumg@rz.uni-potsdam.de}  \\
 {\footnotesize FAX: +49-331-977-1299}    \\[1mm]                 
\and
 {\sc Hendrik Grundling}                                            \\[1mm] 
 {\footnotesize Department of Mathematics,}                         \\ 
 {\footnotesize University of New South Wales,}                      \\
 {\footnotesize Sydney, NSW 2052, Australia.}                             \\ 
 {\footnotesize hendrik@maths.unsw.edu.au}                  \\ 
 {\footnotesize FAX: +61-2-93857123}}

\date{RUNNING TITLE: Superselection and Constraints.}

\begin{document}
\maketitle
\begin{abstract}
For systems which contain both 
superselection structure and constraints, we study compatibility
between constraining and superselection. Specifically, we start with a
generalisation of Doplicher-Roberts superselection theory
to the case of nontrivial centre, and a set of Dirac quantum constraints
and find conditions under which the superselection structures will
survive constraining in some form. This involves an analysis
of the restriction and factorisation of superselection structures.
We develop an example for this theory, modelled on interacting QED.
\end{abstract}
\section{Introduction}

In heuristic quantum field theory, there are many examples of systems
which contain  global charges (hence superselection structure) as well
as a local gauge symmetry (hence constraints). Most of these systems cannot
currently be written in a consistent mathematical framework, due to
the presence of interactions. Nevertheless, the mathematical structure
of superselection by itself has been properly developed 
(cf.~\cite{DR, Bg, BW}), 
as well as the
mathematical structure of quantum constraints \cite{Grundling85, Lledo,
lands}, hence one can at
least abstractly consider systems which contain both.
This will be the focus of our investigations in this paper.
We will address the natural intertwining questions for the
two structures, as well as compatibility issues.

There is a choice in how the problem of superselection with
constraints is posed mathematically. We will be guided by the most
important physics example in this class, which is that of
a quantized local gauge field, acting on a fermion field. 
It has a Gauss law constraint
(implementing the local gauge transformations) as well as
a set of global charges (leading to superselection).

The architecture of the paper is as follows, in Sect.~2 we give a
brief summary of the superselection theory which we intend to use here.
We include recent results concerning the case of an observable algebra
with nontrivial centre (cf.~\cite{BL}), and some new results on
the field algebra.
In Sect.~3 we give a summary of quantum constraints, and
in Sect.~4 we collect our main results. The proofs for these
are in Sect.~6, and in Sect.~5 we present an example.

\section{Fundamentals of superselection}  

In this section  we summarize the structures from superselection
theory which we need. For proofs, we refer to the literature
if possible.

The superselection problem in algebraic quantum field theory, 
as stated by the
Doplicher--Haag--Roberts~(DHR) selection criterion,
led to a profound body of work, culminating in
the general Doplicher--Roberts~(DR) duality theory for compact groups.
The DHR criterion selects a distinguished class of ``admissible''
representations of a quasilocal algebra $\al A.$ of observables,
where the centre
is trivial, i.e.
$Z(\al A.)=\C\un,$
or even
$\al A.$
is assumed to be simple.
This 
corresponds to the selection of a DR--category
$\al T.$
of ``admissible'' endomorphisms
of $\al A.\,.$ 
Furthermore, from this endomorphism category 
$\al T.$
the
DR--analysis constructs a C*--algebra $\al F.\supset\al A.$
together with a compact group action 
$\alpha:\al G.\ni g\to\alpha_{g}\in\aut\al F.$
such that 
\begin{itemize}
\item{}
$\al A.$ is the fixed point algebra of this action,
\item{}
$\al T.$ coincides with the category of all ``canonical
endomorphisms" of $\al A.$ (cf. Subsection~\ref{CanEnd}).
\end{itemize}

$\al F.$  
is called a Hilbert extension of $\al A.$ in \cite{BW}.
Physically, $\al F.$ is identified as a field algebra and
$\al G.$ with a global gauge group of the system.
$\{\al F.,\alpha_{\al G.}\}$
is uniquely determined by
$\al T.$
up to
$\al A.$-module isomorphisms. Conversely,
$\{\al F.,\alpha_{\al G.}\}$
determines uniquely its category of all canonical endomorphisms.
Therefore one can state the equivalence of the ``selection
principle", given by
$\al T.$
and the ``symmetry principle", given by
$\al G.\,$. This duality is one of the crucial theorems of the
Doplicher-Roberts theory.

In contrast to the original theory of Doplicher and Roberts, we allow
here a nontrivial centre for $\al A..$ The reason for this is that when there
are constraints present, the system contains nonphysical information, so
there is no physical reason why $\al A.$ should be simple.
Only after eliminating the constraints should one  require the
final observable algebra to be simple, hence having trivial centre.
Now a duality theorem for a C*--algebra with nontrivial centre has been
proven recently~\cite{BL, B2}, establishing a bijection between
 distinguished 
categories
of endomorphisms of 
$\al A.$ 
and  Hilbert extensions of $\al A.$ satisfying some additional conditions,
of which the most important
is: ${\al A.'\cap\al F.=Z(\al A.)}$
(i.e. the relative commutant is assumed to be minimal).
This will be properly explained below. This condition has already been
used by Mack and Schomerus~\cite{MS}  as a ``new principle".

\subsection{Basic properties of Hilbert systems}

Below $\al F.$ will always denote a unital C*--algebra. 
A Hilbert space $\al H.\subset
\al F.$ is called {\bf algebraic} if the scalar product $\langle\cdot,\cdot
\rangle$ of $\al H.$ is given by
$\langle A,B\rangle\un := A^{\ast}B$ for $A,\; B\in\al H.\,.$ Henceforth
we consider only finite-dimensional algebraic Hilbert spaces. The 
 support 
$\hbox{supp}\,\al H.$
of $\al H.$ is defined by
$\hbox{supp}\,\al H.:=\sum_{j=1}^{d}\Phi_j\Phi_{j}^{\ast}$
where $\{\Phi_j\,\big|\,
j=1,\ldots,\,d\}$ is any orthonormal basis of $\al H..$ 
{\it Unless otherwise specified, we assume below that
each algebraic Hilbert space $\al H.$ considered,
satisfies  ${\rm supp}\,\al H.
=\un.$}

We also fix a compact 
C*-dynamical system
$\{\al F.,\al G.,\alpha\}$,
i.e.
$\al G.$
is a compact group and
$\alpha:\al G.\ni g\to\alpha_{g}\in\aut\al F.$
is a pointwise norm-continuous morphism.
For $\gamma\in\wh{\al G.}$
(the dual of $\al G.$) its {\bf spectral projection}
$\Pi_{\gamma}\in\al L.(\al F.)$
is defined by
\begin{eqnarray*}
\Pi_\gamma(F)&:=&\int_{\al G.}\ol\chi_\gamma(g).\,\alpha_{g}(F)\,dg
\quad\hbox{for all}\quad F\in\al F.,  \\[1mm]
\hbox{where:}\quad\qquad
\chi_\gamma(g)&:=&\dim\gamma\cdot\tr\pi(g),\quad\pi\in\gamma\,
\end{eqnarray*}
and its {\bf spectral subspace} $\Pi_\gamma\al F.$ satisfies
$\Pi_\gamma\al F.=\csp\{\al L.\subset\al F.\}$ where $\al L.$
runs through all invariant subspaces of $\al F.$ 
which transform under $\alpha\s{\al G.}.$ according to $\gamma$
(cf.~\cite{ES}).
Define the {\bf spectrum} of 
$\alpha_{\al G.}$ by
\[
\spec\alpha_{\al G.}:=\set \gamma\in\wh{\al G.}, \Pi_\gamma\not=0.
\]
 
Our central object of study is:

\begin{defi}
The C*-dynamical system
$\{\al F.,\al G.,\alpha\}$
is called a {\bf Hilbert system} if
for each $\gamma\in\wh{\al G.}$
there is an algebraic Hilbert space 
$\al H._\gamma\subset\al F.,$
such that
$\alpha_{\al G.}$
acts invariantly on
$\al H._\gamma,$  
and the unitary representation
$\al G.\rest\al H._\gamma$ 
is in the equivalence class 
$\gamma\in\wh{\al G.}$.
\end{defi}
\begin{rem}
Note that for a Hilbert system $\{\al F.,\al G.,\alpha\}$
we have necessarily that the algebraic Hilbert spaces
satisfy $\al H._\gamma\subset\Pi_\gamma\al F.$ for all $\gamma,$
and hence that $\spec\alpha\s{\al G.}.=\wh{\al G.}$ i.e.
the spectrum is {\it full}.
The morphism
$\alpha:\al G.\to \hbox{Aut}\,\al F.$
is necessarily faithful. So, since
$\al G.$
is compact and
$\hbox{Aut}\,\al F.$
is Hausdorff w.r.t. the topology of pointwise norm-convergence,
$\alpha$
is a homeomorphism of
$\al G.$
onto its image. Thus
$\al G.$
and
$\alpha_{\al G.}$
are isomorphic as topological groups.
\end{rem}

We are mainly interested in Hilbert systems whose fixed point
algebras coincide such that they appear as extensions of it.

\begin{defi}
A Hilbert system 
$\{\al F.,\al G.,\alpha\}$ 
is called a {\bf Hilbert
extension} of a C*--algebra $\al A.\subset\al F.$ if $\al A.$ is the
fixed point algebra of ${\al G.}.$
Two Hilbert extensions 
$\{\al F._i,\,\al G.\,,\alpha^{i}\},\;i=1,\,2$
of $\al A.$
(w.r.t. the same group $\al G.$)
are called 
$\al A.\hbox{\bf--module isomorphic}$ 
if there is an isomorphism 
$\tau:\al F._1\to\al F._2$ 
such that
$\tau(A)=A$ 
for 
$A\in\al A.,$ 
and 
$\tau$ 
intertwines the group actions, i.e.
$\tau\circ\alpha^{1}_g=\alpha^{2}_g\circ\tau.$
\end{defi}

\begin{rem}
\begin{itemize}
\item[(i)]
Group automorphisms of
$\al G.$
lead to $\al A.$-module isomorphic Hilbert extensions of
$\al A.$,
i.e. if
$\{\al F.,\al G.,\alpha\}$
is a Hilbert extension of
$\al A.$
and
$\xi$
an automorphism of
$\al G.$,
then the Hilbert extensions
$\{\al F.,\al G.,\alpha\}$
and
$\{\al F.,\al G.,\alpha\circ\xi\}$
are $\al A.$-module isomorphic.
So the Hilbert system
$\{\al F.,\al G.,\alpha\}$
depends, up to $\al A.$-module isomorphisms, only on
$\alpha_{\al G.}$,
which is isomorphic to
$\al G.$.
In other words, up to $\al A.$-module isomorphism we may identify
$\al G.$
and
$\alpha_{\al G.}\subset\aut\al F.$
neglecting the action
$\alpha$
which has no relevance from this point of view. Therefore in the
following, unless it is otherwise specified, we use the notation
$\{\al F.,\al G.\}$
for a Hilbert extension of
$\al A.$,
where
$\al G.\subset\aut\al F.$.
\item[(ii)]
As mentioned above, examples of
Hilbert systems arise in DHR--superselection theory
cf.~\cite{BW, Bg}.
There are also constructions by means of
tensor products 
of Cuntz algebras
(cf.~\cite{DR2}). In these examples the 
relative commutant
of the fixed point algebra 
$\al A.$,
hence also its center,
is trivial. Another 
construction for
$\al G.=\T$,
by means of the loop group
$C^{\infty}(S^{1},\,\T)$
is in \cite{BC}, and for this
$Z(\al A.)$
is nontrivial.
\end{itemize}
\end{rem}

\begin{rem}
\label{remark1}
A Hilbert system $\{\al F.,\,\al G.\}$ is a highly structured object;-
we list some important facts and properties (for details, consult
~\cite{Bg,BW}):
\begin{itemize}
\item[(i)]
The spectral projections  
satisfy:
\begin{eqnarray*}
\Pi_{\gamma_1}\Pi_{\gamma_2} &=&
\Pi_{\gamma_2}\Pi_{\gamma_1}=\delta\s\gamma_1\gamma_2.\Pi_{\gamma_1}  \\[1mm]
\|\Pi_\gamma\| &\leq& d(\gamma)^{3/2}\;,
\qquad d(\gamma):=\dim(\al H._\gamma)\;,      \\[1mm]
\Pi_\gamma\al F.  &=&  \spa(\al AH._\gamma)\;,    
\quad \Pi_{\iota}\al F.=\al A.\;,
\end{eqnarray*}
where
$\iota\in\wh{\al G.}$
denotes the trivial representation of $\al G..$
\item[(ii)]
 Each $F\in\al F.$ is uniquely determined by its projections 
$\Pi_\gamma F,$ $\gamma\in\wh{\al G.},$ i.e. $F=0$
iff $\Pi_\gamma F =0$ for all $\gamma\in\wh{\al G.},$ 
cf.~Corollary~2.6 of \cite {B2}.
\item[(iii)]
A useful
*-subalgebra of
$\al F.$ is
\[
\al F._{\rm fin}:= \set F\in\al F.,\Pi_\gamma F\not=0\quad\hbox{for
only finitely many $\gamma\in\wh{\al G.}$}.
\]
which is dense in $\al F.$ w.r.t. the C*--norm (cf.~\cite{S}).
\item[(iv)]
In $\al F.$ there is an $\al A.\hbox{--scalar product}$ given by
${\langle F,\, G\rangle_{\al A.}:=\Pi_\iota FG^*},$ w.r.t. which
the spectral projections are symmetric, i.e.
$\bra\Pi_\gamma F,G.=\bra F,\Pi_\gamma G.$ 
for all $F,\; G\in\al F.,$ $\gamma\in\wh{\al G.}$.
Using the $\al A.$-scalar product one can define a norm on
$\al F.,$
called the $\al A.\hbox{-norm}$ 
\[
\vert F\vert_{\al A.}:=\Vert\langle F,F\rangle\s{\al A.}.\Vert^{1/2},\quad
F\in \al F..
\]
Note that
$\vert F\vert_{\al A.}\leq \Vert F\Vert$
and that
$\al F.$
in general is not closed w.r.t. the $\al A.$-norm.
Then
for each $F\in\al F.$ we have that
$F=\sum_{\gamma\in\wh{\al G.}}\Pi_\gamma F$
where the sum on the right hand side is convergent w.r.t. the 
$\al A.\hbox{--norm}$ but not necessarily w.r.t. the C*--norm $\|\cdot\|\,.$
We also have Parseval's equation:
$\langle F,F\rangle_{\al A.}
=\sum_{\gamma\in\wh{\al G.}}\langle\Pi_\gamma F,\Pi_\gamma F\rangle_{\al A.}
\;,$ cf. Proposition~2.5 in \cite{B2}.
Moreover
 $\big|\Pi_\gamma\big|_{\al A.}=1$
for all $\gamma\in\wh{\al G.}$,
where $\vert\cdot\vert_{\al A.}$
denotes the operator norm of $\Pi_\gamma$ w.r.t. the norm
$\vert\cdot\vert_{\al A.}$ in $\al F.$.
\item[(v)]
Generally for a Hilbert system, the assignment $\gamma\to\al H._\gamma$
is not unique.
If
$U\in \al A.$
is unitary then also
$U\al H._\gamma\subset\Pi_{\gamma}\al F.$
is an $\al G.$-invariant algebraic Hilbert space carrying the representation
$\gamma\in\wh{\al G.}.$ Each 
$\al G.$-invariant algebraic
Hilbert space 
$\al K.$
which carries the representation
$\gamma\,$
is of this form, i.e. there is a unitary
$V\in\al A.$
such that
$\al K.=V\al H._{\gamma}.$
For a general $\al G.\hbox{--invariant}$
algebraic Hilbert space $\al H.\subset\al F.,$ 
we may have that $\al G.\rest\al H.$ is not irreducible,
i.e. it need not be of the form $\al K.=V\al H._{\gamma}.$
Below we will consider further conditions on the Hilbert system
to control the structure of these.
\item[(vi)]
Given two $\al G.\hbox{--invariant}$
algebraic Hilbert spaces 
$\al H.,\al K.\subset\al F.,$
then
$\spa(\al H.\cdot\al K.)$
is also a $\al G.\hbox{--invariant}$
algebraic Hilbert space 
which we will 
briefly denote by $\al H.\cdot\al K..$
It is a realization of the tensor product
$\al H.\otimes\al K.$
within
$\al F.$
and
carries the tensor product of the representations of $\al G.$
carried by $\al H.$ and $\al K.$ in the obvious way.
\item[(vii)]
Let
$\al H.,\al K.$ be two $\al G.\hbox{--invariant}$
algebraic Hilbert spaces,
 but not necessarily of support 1. Then there
is a natural isometric embedding 
$\al J.:\al L.(\al H.,\al K.)\to\al F.$
given by
\[
\al J.(T):=\sum_{j,k}
t_{j,k}\Psi_{j}\Phi^{\ast}_{k},\quad t_{j,k}\in \C,
\quad T\in\al L.(\al H.,\al K.)
\]
where
$\{\Phi_{k}\}_{k},\{\Psi_{j}\}_{j}$
are orthonormal bases of
$\al H.$ and $\al K.$ respectively, and where
\[
T(\Phi_{k})=\sum_{j}t_{j,k}\Psi_{j},
\]
i.e.
$(t_{j,k})$
is the matrix of $T$ w.r.t. these orthonormal bases. One has
\[
T(\Phi)=\al J.(T)\cdot\Phi,\quad \Phi\in\al H..
\]
This implies: if
$T_{j}\in\al L.(\al H._{j},\al K._{j}),j=1,2,$
hence
$T_{1}\otimes T_{2}\in\al L.(\al H._{1}\al H._{2},\al K._{1}\al K._{2}),$
then
${\al J.(T_{1}\otimes T_{2})\Phi_{1}\Phi_{2}}=\al J.(T_{1})\Phi_{1}\al J.(T_{2})\Phi_{2}$
for
$\Phi_{j}\in\al H._{j}$.

Moreover
$\al J.(T)\in\al A.$
iff
$T\in\al L._{\al G.}(\al H.,\al K.),$
where
$\al L._{\al G.}(\al H.,\al K.)$
denotes the linear subspace of
$\al L.(\al H.,\al K.)$
consisting of all intertwining operators of the representations
of 
$\al G.$
on 
$\al H.$ and $\al K.$ (cf. p. 222 ~\cite{BW}).

\end{itemize}
\end{rem}

\subsection{The category of $\al G.$-invariant algebraic
Hilbert spaces}

The $\al G.$-invariant algebraic Hilbert spaces
$\al H.$
of
$\{\al F.,\al G.\}$
form the objects of a category
$\al T._{\al G.}$
associated to
$\{\al F.,\al G.\}$
whose arrows are given by the elements of
$(\al H.,\,\al K.):=\al J.(\al L._{\al G.}(\al H.,\al K.))\subset\al A.$.
It is already large enough to carry all tensor products of the representations of
$\al G.$ on its objects by Remark~\ref{remark1}(vi)
(though not necessarily subrepresentations and direct sums).
First, let us state 
some 
of its rich structure (cf.~\cite{BW,DR2}):
\begin{pro} 
\label{perm&conj}
For $\{\al F.,\al G.\}$ the category $\al T._{\al G.}$
is a {\bf tensor C*-category}, i.e. the arrow spaces $(\al H.,\,\al K.)$
are Banach spaces such that \chop
$\bullet$ w.r.t. composition of arrows $R,\, S$ we have
$\|R\circ S\|\leq\|R\|\|S\|,$\chop
$\bullet$ there is an antilinear involutive contravariant functor
$*:\al T._{\al G.}\to\al T._{\al G.}$ such that
$\|R^*\circ R\|=\|R\|^2$ for all arrows $R$ with the same range and domain
(here the functor $*$ is given by the involution in $\al F.)$\chop
$\bullet$ There is an associative product $\cdot$ on $\ob\al T._{\al G.}$ 
and an identity object $1\in\ob\al T._{\al G.}$ (i.e. $1\cdot\al H.=\al H.
=\al H.\cdot 1)$ and there is an associative bilinear product $\times$
of the arrows, such that if $R\in(\al H.,\al K.)$ and
$R'\in(\al H.',\al K.')$ then $R\times R'\in(\al H.\cdot\al H.',\,
\al K.\cdot\al K.').$ Moreover we require that for $R,\;R'$ as above:
\begin{equation}
  1_\iota\times R=R\times 1_\iota=R\;,\quad (R\times R')^*=R^*\times R'^*\,,
\end{equation}
where $1_\iota\in(1,1)$ is the identity arrow, as well as the interchange law
\[
 (S\circ R)\times (S'\circ R')=(S\times S')\circ (R\times R')\,,
\]
whenever the left hand side is defined.\chop
Here in $\al T._{\al G.},$ the product $\cdot$ is given by the product
of $\al F.,$ 
the identity object is
$1:=\C\un$ and the product
$\times$ is defined by
\[
R\times R':= \al J.(T\otimes T'),
\]
for
$R=\al J.(T),\,R'=\al J.(T'),$ where $T\in\al L._{\al G.}(\al H.,\al K.),\,
T'\in\al L._{\al G.}(\al H.',\al K.').$
Note that
$(1,1)=(\C\un,\C\un)=\C\un$,
i.e.
$1_{\iota}=\un.$
\end{pro}
$\al T._{\al  G.}$ has additional important structures
(permutation and conjugation), which we will consider below in
Subsection~\ref{PermConj}.

We need to examine conditions to require of $\{\al F.,\,
\al G.\}$ to ensure that $\al T._{\al G.}$
carries subrepresentations and direct sums.

\begin{defi}
Let $\al H.,\,\al K.\in\ob\al T._{\al G.},$ and define
$\al H.<\al K.$
to mean that there is an orthoprojection
$E$
on
$\al K.$
such that
$E\al K.$
is invariant w.r.t.
$\al G.$
and the representation
$\al G.\rest\al H.$
is unitarily equivalent to
$\al G.\rest E\al K..$
Call $\al H.$ a {\bf subobject} of $\al K.\,.$
\end{defi}

It is easy to see that $<$ is a partial order.
Note that
$\al H.<\al K.$
iff there is an isometry
$V\in\al L._{\al G.}(\al H.,\al K.)$
such that
$VV^{\ast}=:E$
is a projection of
$\al K.$,
i.e.
$V\al H.=E\al K..$
Then
$\al J.(V)\in\al A.$
and
$E\al K.=\al J.(V)\cdot\al H..$

If
$E\in\al L._{\al G.}(\al K.)$
is an orthoprojection
$0<E<\un,$
i.e.
$E$
is a reducing projection for the representation of
$\al G.$
on
$\al K.$,
then the question arises whether there is an object
$\al H.$
such that the representations on
$\al H.$
and
$E\al K.$
are unitarily equivalent. This suggests the concept of {\it
closedness} of
$\al T._{\al G.}$
w.r.t. subobjects.

\begin{defi}
The category
$\al T._{\al G.}$
is {\bf closed w.r.t. subobjects} if to each
$\al K.\in\hbox{Ob}\,\al T._{\al G.}$
and to each nontrivial orthoprojection
$E\in\al L._{\al G.}(\al K.)$
there is an isometry
$\wh{V}\in\al A.$
with
$\wh{V}\wh{V}^{\ast}=\al J.(E).$
In this case
$\al H.:=\wh{V}^{\ast}\cdot\al K.$
is a subobject
$\al H.<\al K.$
assigned to
$E$,
where
$\wh{V}=\al J.(V)$ for some isometry $V\in\al L._{\al G.}(\al H.,\al K.)$
with $VV^*=E.$
\end{defi}

Next, we consider when an object of $\al T._{\al G.}$ carries
the direct sum of the representations of two other objects.
If
$V,W\in\al A.$
are isometries with
$VV^{\ast}+WW^{\ast}=\un$
and
$\al H.,\al K.\in\hbox{Ob}\,\al T._{\al G.}$
then 
we call the algebraic Hilbert space
$V\al H.+W\al K.$
of support 1 a {\bf direct sum} of
$\al H.$ and $\al K.$.
It is $\al G.$-invariant and carries the direct sum of the
representations on
$\al H.$ and $\al K.$
but in general  depends on the choice of isometries $V,\;W.$
We define

\begin{defi}
\begin{itemize} 
\item[(i)]
The category
$\al T._{\al G.}$
is {\bf closed w.r.t. direct sums} if to each
$\al H._{1},\al H._{2}\in\ob\,\al T._{\al G.}$
there is an object
$\al K.\in\ob\,\al T._{\al G.}$
and there are isometries
$V_{1},V_{2}\in\al A.$
with
$V_{1}V_{1}^{\ast}+V_{2}V_{2}^{\ast}=\un$
such that
$\al K.=V_{1}\al H._{1}+V_{2}\al H._{2}$
(then
$V_{1}\in(\al H._{1},\al K.)$
and
$V_{2}\in(\al H._{2},\al K.)$
follow).
\item[(ii)]
A C*-algebra
$\al A.$
satisfies {\bf Property B} if there are isometries
$V_{1},V_{2}\in\al A.$
such that
$V_{1}V_{1}^{\ast}+V_{2}V_{2}^{\ast}=\un.$
A Hilbert system
$\{\al F.,\al G.\}$
is said to satisfy {\bf Property B} if its fixed point algebra
$\al A.:=\Pi_{\iota}\al F.$
satisfies Property B.
\end{itemize}
\end{defi}

\begin{rem}
For a Hilbert system
$\{\al F.,\al G.\}$ we have:
\begin{itemize}
\item[(i)] It
satisfies Property B iff
$\al T._{\al G.}$
is closed w.r.t. direct sums.
\item[(ii)]
For nonabelian
$\al G.,$
the category $\al T._{\al G.}$ is closed w.r.t. subobjects iff
it is closed w.r.t. direct sums iff it has Property B 
 cf.~Prop.~3.5 of~\cite{B2}.
\item[(iii)]
In the case that
$\al G.$
is abelian, the theory simplifies.
This is because we already have Pontryagin's duality theorem, 
hence it is not necessary to consider closure under
subobjects and direct sums to obtain a duality theory.
\end{itemize}
\end{rem}

\subsection{The category of canonical endomorphisms}
\label{CanEnd}

The main aim of DR--theory is to obtain an intrinsic structure on $\al A.$
from which we can reconstruct the Hilbert system $\HS$ in an essentially unique
way. Here we want to transport the rich structure of $\al T._{\al G.}$ to $\al A.\,.$
\begin{defi}
To each $\al G.\hbox{--invariant}$ algebraic Hilbert space $\al H.\subset
\al F.$  
there is assigned a corresponding 
{\bf inner endomorphism} $\rho\s{\al H.}.\in\endo\al F.$ given by
\[
\rho\s{\al H.}.(F):=\sum_{j=1}^{d(\al H.)}\Phi_jF\Phi_j^*\;,
\]
where $\{\Phi_j\,\big|\,j=1,\ldots,\,d(\al H.)\}$ is any orthonormal basis of
$\al H..$ Note that $\rho\s{\al H.}.$ preserves $\al A.\,.$
A {\bf canonical endomorphism} is the restriction of an inner endomorphism
to $\al A.,$ i.e. it is of the form 
$\rho_{\al H.}\rest\al A.\in\mbox{End}\,\al A.\,.$
\end{defi}  

\begin{rem}
\label{remark2}
\begin{itemize}
\item[(i)]
The definition  of the canonical endomorphisms
uses
$\al F.$
explicitly. The question arises whether the 
canonical
endomorphisms can be
characterised by intrinsic properties 
within
$\al A.$. 
This interplay between the
$\rho_{\al H.}$
and the
$\rho_{\al H.}\rest\al A.$
plays an essential role in the DR-theory. 
Below, we omit
the restriction symbol and regard the
$\rho_{\al H.}$
also as endomorphisms of
$\al A.$.
\item[(ii)]
If the emphasis is only on the representation
$\gamma$
and not on its corresponding algebraic Hilbert space
$\al H._{\gamma},$
 we will write
$\rho_{\gamma}$
instead of
$\rho_{\al H._{\gamma}}$.
\item[(iii)]
Note that $\Phi A=\rho\s{\al H.}.(A)\Phi$ for all 
$\Phi\in\al H.$ and $A\in\al A..$
\item[(iv)]
Note that the identity endomorphism $\iota$ is assigned to
$\al H.=\C\un,$
i.e. $\rho\s{\C\un}.:=\iota.$
\item[(v)]
Let
$\al H.,\al K.$
be as before, then 
$\rho\s{\al H.}.\circ\rho\s{\al K.}.=\rho\s{\al H.\cdot\al K.}..$
\item[(vi)]
The map $\rho$ from $\ob\al T._{\al G.}$ to the canonical endomorphisms
is in general not injective. In fact we have:
if $\al H.,\al K.\in\ob\al T._{\al G.},$
then
$\rho_{\al H.}\rest\al A.=\rho_{\al K.}\rest\al A.$
iff
$\Psi^{\ast}\Phi\in\al A.'\cap\al F.$
for all
$\Phi\in\al H.,\Psi\in\al K.\,,$
cf. Prop.~3.9 in \cite{B2}.
\end{itemize}
\end{rem}

\begin{defi}
Define $\al T.$ to be the category with objects the 
canonical endomorphisms, and arrows the intertwiner spaces, where the
{\bf intertwiner space} of canonical endomorphisms 
$\sigma,\;\tau\in\endo\al A.$ 
is:
\[
(\sigma,\,\tau):=\set X\in\al A.,X\sigma(A)=\tau(A)X\quad\hbox{for all}
\;{A\in\al A.}.\;.
\]
and this is a complex Banach  space.
For $A\in(\sigma,\sigma'),$ $B\in(\tau,\tau'),$ we define
${A\times B}:=A\sigma(B)\in{(\sigma\tau,\,\sigma'\tau')\,.}$
We will say that $\sigma,\;\tau\in\endo\al A.$ are
{\bf mutually disjoint} if  ${(\sigma,\,\tau)}=\{0\}$ when
$\sigma\not=\tau.$
\end{defi}

\begin{rem}
\label{remark3}
\begin{itemize}
\item[(i)]
We have
 $(\iota,\,\iota)=Z(\al A.):=$ centre of $\al A..$ 
 \item[(ii)]
 The composition of two canonical endomorphisms (which corresponds to
products of the generating Hilbert spaces, see Remark 2.13 (v), i.e. to
 tensor products of representations) satisfies the correct compatibility conditions
 with the product $\times$ of intertwiners
 to ensure that $\al T.$ is a  C*-tensor category cf. Prop.~\ref{perm&conj} and
 \cite{DR}.
The identity object is $\iota.$
\item[(iii)] Recall the isometry 
$\al J.:\al L.\s{\al G.}.(\al H.,\al K.)
\longrightarrow \al A.$ encountered in Remark~\ref{remark1}(vii). 
We claim that
its image is in fact contained in ${\left(\rho\s{\al H.}.,\,
\rho\s{\al K.}.\right)}.$ To see this, let $\Phi\in\al H.,\;
A\in\al A.$ 
and
$T\in\al L._{\al G.}(\al H.,\al K.).$
Then
\[
\al J.(T)\rho\s{\al H.}.(A)\Phi=\al J.(T)\Phi\cdot A=
T(\Phi)\cdot A=\rho\s{\al K.}.(A)T(\Phi)=
\rho\s{\al K.}.(A)\al J.(T)\cdot\Phi
\]
hence 
\[
\al J.(T)\rho\s{\al H.}.(A)=\rho\s{\al K.}.(A)\al J.(T)
\]
i.e. $\al J.(T)\in {\left(\rho\s{\al H.}.,\,
\rho\s{\al K.}.\right)}$
or
\begin{equation}
\label{nonful}
(\al H.,\,\al K.)=
\al J.(\al L._{\al G.}(\al H.,\al K.))\subseteq
(\rho_{\al H.},\,\rho_{\al K.}).
\end{equation}
In general, the inclusion is proper.
Note that for
$R\in(\al H.,\al K.),\; R'\in(\al H.',\al K.')$
we have
$\al J.(R\otimes R')=R\rho_{\al H.}(R'),$
i.e.
$\times$
restricted to the
$(\al H.,\al K.)'s$
coincides with the definition of
$\times$
in Proposition~\ref{perm&conj} of the category
$\al T._{\al G.}.$
\end{itemize}
\end{rem}
 
Next we would like to define the concepts of subobject and direct sums
on $\ob\al T.$  compatibly with those on $\ob\al T._{\al G.}$
under the morphism $\rho.$ Recall that $\al H.<\al K.$ iff we have an isometry
$V\in\al L.\s{\al G.}.(\al H.,\,\al K.)$ and a projection $E\in
\al L.\s{\al G.}.(\al K.)$
with $V\al H.=E\al K.=\al J.(E)\cdot\al K.=\al J.(V)\cdot\al H..$
Then by (\ref{nonful}) we get that     
 $\al J.(V)\in
 (\rho_{\al H.},\rho_{\al K.})$
 and
 $\al J.(E)\in
 (\rho_{\al K.},\rho_{\al K.}).$

Note that if $\al L.=V\al H.+W\al K.$
for isometries
$V,\,W\in\al A.$ with $VV^*+WW^*=\un,$ then
$V\in(\rho\s{\al H.}.,\rho\s{\al L.}.)$ and $W\in(\rho\s{\al K.}.,
\rho\s{\al L.}.).$

\begin{defi}
\begin{itemize}
\item[(i)]
$\tau\in\ob\al T.$ is a {\bf subobject} 
 of $\sigma\in\ob\al T.,$
denoted ${\tau<\sigma,}$ if
there 
is an isometry
$V\in (\tau,\sigma)$.
In this case
$\tau(\cdot)=V^{\ast}\sigma(\cdot)V$
and
$VV^{\ast}=:E\in (\sigma,\sigma)$
follow.
 
\item[(ii)]
$\lambda\in\ob\,\al T.$
is a {\bf direct sum} of
$\sigma,\tau\in\ob\,\al T.,$
if there are isometries
$V\in(\sigma,\lambda),\,W\in(\tau,\lambda)$
with
$VV^{\ast}+WW^{\ast}=\un$
such that
\[
\lambda(\cdot)=V\sigma(\cdot)V^{\ast}+W\tau(\cdot)W^{\ast}.
\]
\end{itemize}
\end{defi}

\begin{rem}
\label{Tsubobj}
\begin{itemize}
\item[(i)]
The subobject relation
$\tau<\sigma$
is again a partial order, because
$\tau<\sigma$
and
$\sigma<\mu$
imply the existence of isometries
$V\in (\tau,\sigma),\,W\in (\sigma,\mu)$.
Then
$WV\in (\tau,\mu)$
is also an isometry, i.e.
$\tau<\mu$.
\item[(ii)] 
A direct sum as defined above
is only unique up to unitary equivalence, i.e. if
$\lambda,\,\lambda'$
are direct sums of  $\sigma,\tau\in\ob\,\al T.,$ then there is a unitary
$U\in(\lambda,\lambda').$
\item[(iii)]
We have
$\rho\s{V\al H.+W\al K.}.(\cdot)=V\rho_{\al H.}(\cdot)V^{\ast}+
W\rho_{\al K.}(\cdot)W^{\ast}$
where
the isometries $V,\,W\in\al A.$ satisfy $VV^*+WW^*=\un\,.$
Also, if $\al H.<\al K.,$ then
$\tau:=\rho_{\al H.}<\rho_{\al K.}=:\sigma.$
However, this does not mean that the partial order
$\tau<\sigma$
can be {\it defined} by
$\al H.<\al K.$
because
the transitivity can be violated for some choices of $\al H.,\,\al K.$
cf. Remark~\ref{remark2}(vi).

\end{itemize}
\end{rem}

The closedness of
$\al T.$
w.r.t. direct sums is defined by the closedness of
$\al T._{\al G.}$
w.r.t. direct sums.
The closedness w.r.t. subobjects for
$\al T.$
is defined by the closedness w.r.t. subobjects for
$\al T._{\al G.}$
in the following sense: If
\begin{equation}
\label{endoH}
\lambda=\rho_{\al H.}\in\ob\al T.
\end{equation}
is given then for all
$\al H.$
satisfying (\ref{endoH}) and to each nontrivial
projection
$E\in\al J.(\al L._{\al G.}(\al H.))$
there is an isometry
$V\in\al A.$
with
$VV^{\ast}=E.$
Then

\begin{pro}
If
$\{\al F.,\al G.\}$ is nonabelian and
satisfies Property B then
$\al T.$
is closed w.r.t. direct sums and subobjects.
\end{pro}

\subsection{Connection between $\al T._{\al G.}$ and $\al T.$
and further structures.}
\label{PermConj}

In the following we assume that
$\{\al F.,\al G.\}$
satisfies Property B. 

There is a very important relation between the two categories $\al T._{\al G.}$
and $\al T.,$ obtained as follows. The two assignments
$\rho:\hbox{Ob}\,\al T._{\al G.}\to\al T.$
(by
$\al H.\to\rho_{\al H.})$
and
$\al J.:\al L._{\al G.}((\al H.,\al K.))\to(\rho_{\al H.},\rho_{\al K.})$
combine into a faithful categorial morphism from
$\al T._{\al G.}$
to
$\al T.$
which is compatible with direct sums and subobjects (cf. Remark~\ref{Tsubobj}(iii)) but 
is not full in general, i.e. the inclusion in Equation~(\ref{nonful}) 
is improper for some 
$\al H.$
and
$\al K.$.
If
$\al A.'\cap\al F.=\C\un,$
then this categorial morphism becomes an isomorphism, cf. Prop. 3.12 in \cite{B2}.

The category $\al T._{\al G.}$ 
has the following additional structures
(\cite{BW,DR2}):

\begin{pro} 
\label{perm&conj2}
For $\{\al F.,\al G.\}$ the category $\al T._{\al G.}$
satisfies:
\begin{itemize}
\item[(1)] it has a {\bf permutation structure}, i.e. a map
$\epsilon$ from
$\ob\al T._{\al G.}\times\ob\al T._{\al G.}$
into the arrows such that
\begin{itemize}
\item[(i)]
$\epsilon(\al H.,\al K.)\in (\al H.\al K.,\al K.\al H.)$
is a unitary.
\item[(ii)]
$\epsilon(\al H.,\al K.)\epsilon(\al K.,\al H.)=\un$.
\item[(iii)]
$\epsilon(1,\,\al H.)=\epsilon(\al H.,1)=\un$.
\item[(iv)]
$\epsilon(\al H.\al K.,\al L.)=\epsilon(\al H.,\al L.)\rho_{\al H.}(\epsilon
(\al K.,\al L.))$.
\item[(v)]
$\epsilon(\al H.',\al K.')A\times B=B\times A\epsilon(\al H.,\al
K.)$\quad for all $A\in (\al H.,\al H.'),\,B\in (\al K.,\al K.')$.
\end{itemize}
For $\al T._{\al G.}$ the permutation structure is
 given by
 \[
\epsilon(\al H.,\al K.):=\al J.(\Theta(\al H.,\al K.))
=\sum_{j,k}\Psi_{j}\Phi_{k}\Psi_{j}^{\ast}\Phi_{k}^{\ast}
\]
where $\Theta$ is the flip operator
$\al H.\otimes\al K.\rightarrow\al K.\otimes\al H.,$
and where
$\{\Phi_{k}\}_{k},\,\{\Psi_{j}\}_{j}$
are orthonormal bases of
$\al H.$ and $\al K.$, respectively.
\item[(2)] It has a {\bf conjugation structure}
i.e. for each
$\al H.\in\ob\al T._{\al G.}$
there is a conjugated object
$\overline{\al H.}\in\ob\al T._{\al G.}$,
carrying the conjugated representation of
$\al G.$
and there are  conjugate arrows
$R_{\al H.}\in(1,\overline{\al H.}\al H.),$
$S_{\al H.}=\epsilon(\overline{\al H.},\al H.)R_{\al H.}$
such that
\[
S_{\al H.}^{\ast}\rho\s{\al H.}.(R_{\al H.})=\un,\quad
R_{\al H.}^{\ast}\rho\s{\overline{\al H.}}.(S_{\al H.})=\un.
\]
For $\al T._{\al G.}$ we have
$R_{\al H.}:=\sum_{j}\overline{\Psi}_{j}\Psi_{j}$,
where
$\{\overline{\Psi}_{j}\}_{j}$
is an orthonormal basis of
$\overline{\al H.}$.
If
$\al H.$
carries the representation
$\oplus_{j}\gamma_{j},\,\gamma_{j}\in\wh{\al G.},$
then
$\overline{\al H.}$
is given by a direct sum of
$\al H._{\overline{\gamma_{j}}},$
where
$\overline{\gamma_{j}}\in\wh{\al G.}$
represents the conjugated representation of
$\gamma_{j}.$
\end{itemize}
\end{pro}

\begin{rem}
Using the categorial morphism from
$\al T._{\al G.}$
to
$\al T.$
we equip
$\al T.$
with the image permutation and conjugation structures of those on
$\al T._{\al G.}$.
Note that for the image permutation structure in
$\al T.,$
property (v) need not hold for {\it all} arrows (cf. Remark~\ref{remark3}(iii)).
\end{rem}

For the next definition, observe
first that from the operations defined for an abstract tensor category
(cf. Prop.~\ref{perm&conj}), we can define isometries and projections in
its arrow spaces, i.e. an arrow $V\in(\lambda,\tau)$ is an isometry if
$V^*\circ V=1_\lambda,$ and an arrow $E\in(\lambda,\lambda)$
is a projection if $E=E^*=E\circ E\,.$

\begin{defi}
An {\bf (abstract) DR-category} is an
(abstract) tensor C*-category $\al C.$ 
with $(1,1)=\C\un$ which has a permutation and a conjugation structure,
and has direct sums 
and subobjects, i.e. to all objects $\lambda,\sigma$ there is an object $\tau$
and isometries $V\in(\lambda,\tau),\,W\in(\sigma,\tau)$
such that $VV^{\ast}+WW^{\ast}=1_{\tau}$,
and to each nontrivial projection $E\in(\lambda,\lambda)$ there is an object
$\sigma$ and an isometry $V\in(\sigma,\lambda)$
such that $E=VV^{\ast}.$
\end{defi}

If the Hilbert system $\HS$ satisfies Property B then $\al T._{\al G.}$ 
is an example of a DR-category, but not necessarily $\al T.$
(since property (v) in Prop.~\ref{perm&conj2} need not hold for all arrows).
However, if additionally
$\al A.'\cap\al F.=\C\un$
holds then also
$\al T.$
is a DR-category.
\subsection{Duality Theorems}

Unless otherwise specified, in the following we assume Property
B for $\HS$ when
$\al G.$
is nonabelian.
The DR-theorem 
produces 
 a bijection between pairs
\[ 
    \{\al A.,\al T.\} \quad\hbox{and}\quad \HS\,,
\]
where $\al T.$ is a DR-category of unital endomorphisms of the
unital C*-algebra $\al A.$ with $Z(\al A.)=\C\un,$ and
$\HS$ is a Hilbert extension of $\al A.$ having trivial relative 
commutant, i.e.~$\al A.'\cap\al F.=\C\un$   (see 
\cite{DR,DR5,B3}).
The DR-theorem says that in the case of
Hilbert extensions of $\al A.$
with trivial relative commutant, the category $\al T. $
of all canonical endomorphisms can indeed be characterized
intrinsically by their abstract algebraic properties
as endomorphisms of $\al A.$
and a corresponding bijection can be established.

In this subsection we want to state how to obtain such a bijection
for C*-algebras $\al A.$ with nontrivial center 
$\al Z.\supset\C\un.$
A first problem is that the category $\al T._\al G.$ and
$\al T.$ {\em need not} be isomorphic anymore, cf. Remark~\ref{remark3}(iv)
and Remark~\ref{remark2}(vi), since now we have
\[
 \C\un\not=\al Z.\subseteq\al A.'\cap\al F.\,.
\]
We will investigate in the following the class
of Hilbert extensions $\HS$ with compact group $\al G.$ 
and where the relative commutant satisfies the following 
{\em minimality} condition

\begin{defi}
A Hilbert system
$\{\al F.,\al G.\}$
is called {\bf minimal} if the condition

\begin{equation}
\label{CZ}
\al A.'\cap\al F.=Z(\al A.)\;.
\end{equation}

is satisfied.
\end{defi}
Then we have cf. Prop.~4.3 of \cite{B2}:
\begin{pro}
\label{disjZ}
Let
$\HS$
be a given Hilbert system. Then
$\al A.'\cap\al F.=Z(\al A.)$ iff
$(\rho_{\gamma},\rho_{\gamma'})=\{0\}$
for
$\gamma\neq\gamma'$,
i.e. iff the set
$\{\rho_{\gamma}\;\big|\;\gamma\in\wh{\al G.}\}$
is mutually disjoint.
\end{pro}

Observe that in any Hilbert system,  
for each $\tau\in
\ob\al T.$ the space ${\got h}_\tau:=
\al H._\tau Z(\al A.),$ (where $\al H._\tau $
is a $\al G.\hbox{--invariant}$ algebraic Hilbert space)
is a $\al G.\hbox{--invariant}$ right Hilbert
$Z(\al A.)\hbox{--module}$ i.e. there is a nondegenerate inner product taking
its values in $Z(\al A.)$
and it is $\langle A,B\rangle=A^*B\,.$
Now we have cf. {Prop. 3.1 \cite{BL2}}:

\begin{pro}
\label{prop0}
Let $\HS$ be a given minimal Hilbert system,
then the correspondence ${\tau\leftrightarrow{\got h}_\tau}$ 
is a bijection. Thus ${\got h}_\tau=
\al H._\tau Z(\al A.)$ is independent of the choice of
$\al H._\tau,$ providing that  $\tau=\rho\s{\al H._\tau}..$
This bijection satisfies the conditions
\begin{eqnarray*}
\sigma\circ\tau &\longleftrightarrow&
{\got h}_\sigma\cdot{\got h}_\tau    \\[1mm]
\lambda=(\Ad V)\circ\sigma+(\Ad W)\circ\tau
&\longleftrightarrow&
{\got h}\s\lambda.=V{\got h}_\sigma+W{\got h}_\tau\;.
\end{eqnarray*}
\end{pro}

Thus for minimal Hilbert systems, the 
$Z(\al A.)\hbox{--modules}$ ${\got h}_\tau$
are uniquely determined by their
canonical endomorphisms $\tau,$ even though the choice of
$\al H._{\tau}$ is not unique. We are now interested in those
choices of $\al H._{\tau}$ which are compatible with products:

\begin{defi}
A Hilbert system
$\HS$
is called {\bf regular} if there is an assignment
$\sigma\rightarrow\al H._{\sigma}$
from
$\ob\,\al T.$
to
$\al G.$-invariant algebraic Hilbert spaces in
$\al F.$
such that
\begin{itemize}
\item[(i)] $\sigma=\rho\s{\al H._\sigma}.,$ 
i.e. $\sigma$ is the canonical endomorphism of $\al H._\sigma,$
\item[(ii)] $\sigma\circ\tau\to\al H._\sigma\cdot\al H._\tau\;.$
\end{itemize}
\end{defi}

In a minimal Hilbert system regularity means that there is a
``representing" Hilbert space
$\al H._{\tau}\subset{\got h}_{\tau}$
for each $\tau$ with
${\got h}_{\tau}=\al H._{\tau}Z(\al A.)$
such that the compatibility relation (ii) holds.

If a Hilbert system is minimal and
$Z(\al A.)=\C\un$
then it is necessarily regular.
Thus a class of examples which are trivially minimal and regular,
is provided by DHR--superselection theory.
A nontrivial example of a minimal and regular Hilbert system is constructed in \cite{B2}.

Then we obtain, cf. Theorem 4.9 of \cite{B2}:

\begin{teo}
\label{Teo1}
Let $\HS$ be a minimal and regular Hilbert system, then:
$\al T.$ contains a C*--subcategory $\al T._\C$ with
the same objects, $\ob\al T._\C=\ob\al T.,$ and arrows
$(\sigma,\,\tau)_\C:=(\al H._\sigma,\al H._\tau)=
\al J.\big(\al L.\s{\al G.}.(\al H._\sigma,\al H._\tau)\big)\subset(\sigma,\tau)$ 
such that:
\begin{itemize}
\item[P.1] 
$\quad\al T._{\C}$ is a DR-category (in particular $(\iota,\iota)_{\C}
=\C\un)$.
\item[P.2.] 
$\quad(\sigma,\tau)=(\sigma,\tau)\s\C.\sigma(Z(\al A.))
=\tau(Z(\al A.))(\sigma,\tau)\s\C.\sigma(Z(\al A.))\;.$
\end{itemize}
\end{teo}

\begin{rem}
\begin{itemize}
\item[(i)]
The conditions P.1-P.2 imply that each basis of
$(\sigma,\tau)_{\C}$
is simultaneously a module basis of
$(\sigma,\tau)$
modulo
$\sigma(\al Z.(\al A.))$
as a right module,
i.e. the module
$(\sigma,\tau)$
is free.
\item[(ii)]
We will call the DR-subcategory $\al T._\C$ in Theorem~\ref{Teo1}
{\bf admissible}.
If ``minimality" is omitted 
from the hypotheses of 
Theorem~\ref{Teo1}, then property P.1 remains valid, but not
P.2. In this case  $\al T._\C$ is a DR-subcategory only.
A construction of an example with admissible
subcategory can be found in \cite{B2}.
\end{itemize}
\end{rem}

The converse of Theorem~\ref{Teo1} is also true, and states the main duality result
cf.~\cite{BL}:
\begin{teo}
\label{Teo2}
Let $\al T.$ be a C*--tensor category of unital endomorphisms
of $\al A.$  
and let $\al T.\s\C.$ 
be an admissible (DR-)subcategory. 
Then there is a minimal and regular Hilbert
extension $\HS$ of $\al A.$
 such that $\al T.$ is isomorphic 
to the category of all canonical endomorphisms of $\HS.$
Moreover, if $\al T.\s\C.,\;\al T.'_\C$ are two admissible subcategories of
$\al T.,$ 
then the corresponding Hilbert
extensions are $\al A.\hbox{--module}$ isomorphic iff $\al T.\s\C.$
is {\bf equivalent} to $\al T.'_\C$ i.e. iff there is a map
$V$ from $\ob\al T.$ to the arrows such that:
\begin{eqnarray*}
 V_{\lambda}&\in& (\lambda,\lambda),\quad
V_{\lambda} \quad\hbox{is unitary, and}\quad V_{\lambda\circ\sigma}=V_{\lambda}\times
V_{\sigma}, \\[1mm]
(\lambda,\sigma)_{\C}'&=&V_{\sigma}(\lambda,\sigma)_{\C}V_{\lambda}^{\ast}
\subset (\lambda,\sigma)
\end{eqnarray*}
and we have the following compatibility relations for the corresponding permutators
$\epsilon,\,\epsilon'$ and conjugates $R_\lambda,\,R'_\lambda:$
\begin{eqnarray*}
\epsilon'(\lambda,\sigma)
&=&(V_\sigma\times V_{\lambda})\cdot\epsilon(\lambda,\sigma)\cdot(V_\lambda\times 
V_\sigma)^{\ast} \\[1mm]
R'_\lambda&=&V_{\ol\lambda.\circ
\lambda}R_\lambda,\qquad S'_\lambda
=\epsilon'(\lambda,\ol\lambda.)R'_\lambda.
\end{eqnarray*}
\end{teo}
Thus, in minimal and regular Hilbert systems  there
is an intrinsic characterization of the category of all
canonical endomorphisms in terms of 
$\al A.$
only. Moreover,
up to $\al A.\hbox{--module}$ isomorphisms, there is a bijection
between minimal and regular Hilbert extensions 
and C*-tensor categories $\al T.$ of unital endomorphisms of $\al A.$
with admissible subcategories.  

Note that Theorem~\ref{Teo2} is a generalization of the
DR-theorem for
the case of nontrivial centre
$Z(\al A.)\supset\C\un,$ i.e. it contains the case of the DR-theorem, in that if
$Z(\al A.)=\C\un$
then $\al T.$
itself is admissible (hence a DR-category) and the corresponding
Hilbert extensions have trivial relative commutant.

\subsection{Hilbert systems with abelian groups}
\label{abelianHS}

If
$\al G.$
is abelian the preceding structure simplifies radically. Specifically,
$\wh{\al G.}$
is a discrete abelian group (the character group),
each $\al H._\gamma,\,\gamma\in\wh{\al G.}$ is one--dimensional with a generating
unitary $U_\gamma,$ hence the canonical endomorphisms 
$\rho_{\al H._{\gamma}}$ 
(denoted  by
$\rho_{\gamma},)$
are in fact automorphisms,
necessarily outer on $\al A..$
Since
$\rho_{\gamma_{1}}\circ\rho_{\gamma_{2}}=\rho_{\gamma_{1}\gamma_{2}}$
in this case the set
$\Gamma$
of all canonical endomorphisms
$\rho_{\al H._{\gamma}}$
is a group with the property
$\wh{\al G.}\cong\Gamma/\hbox{int}\,\al A..$
Hence it is not necessary to consider
direct sums, i.e. Property B for
$\al A.$
can be dropped.

In the case
$Z(\al A.)=\C\un$
the permutators $\epsilon$ (restricted to
$\wh{\al G.}\times\wh{\al G.}$)
are elements of the second cohomology group
$H^{2}(\wh{\al G.})$
and
\[
U_{\gamma_{1}}\cdot U_{\gamma_{2}}=\omega(\gamma_{1},\gamma_{2})
U_{\gamma_{1}\circ\gamma_{2}},
\]
where
\[
\epsilon(\gamma_{1},\gamma_{2})=\frac{\omega(\gamma_{1},\gamma_{2})}
{\omega(\gamma_{2},\gamma_{1})}
\]
and $\omega$ is a corresponding 2-cocycle. The field algebra $\al F.$
is just the $\omega\hbox{--twisted}$ discrete crossed product
of $\al A.$ with $\wh{\al G.}$
(see e.g. p.86 ff.~\cite{Bg} for details).
For the case
$Z(\al A.)\supset\C\un$
see ~\cite{BL3} (though the minimal case is not mentioned there).

\section{Kinematics for Quantum Constraints.}
\label{TProcedure}

In this section we give a brief summary of the method of imposing quantum
constraints, developed by Grundling and Hurst \cite{Grundling85,Grundling88b,
Lledo}. There are quite a number of diverse quantum
constraint methods available in the literature at various levels of
rigour (cf.~\cite{lands}). The one we use here is the most congenial from
the point of view of C*--algebraic methods.
Our starting point is:

 \begin{defi}
A {\bf quantum system with constraints} is a
pair $(\al B.,\;\al C.)$ where the {\bf system algebra}
$\al B.$ is a unital {\rm C*}--algebra containing
the {\bf constraint set} $\al C.=\al C.^*.$ A
{\bf constraint condition} on $(\al B.,\,\al C.)$ consists of
the selection of the physical 
state space by:
\[
  {\got S}_D:=\Big\{ \omega\in{\got S}({\al B.})\mid\pi_\omega(C)
              \Omega_\omega=0\quad {\forall}\, C\in {\al C.}\Big\}\,,
\]
where ${\got S}({\al B.})$ denotes the state space of $\al B.$,
and $(\pi_\omega,\al H._\omega,\Omega_\omega)$ denotes the
GNS--data of $\omega$. The elements of ${\got S}_D$ are called 
{\bf Dirac states}.
The case of {\bf unitary constraints} means 
that $\al C.=\al U.-\EINS$, $\;\al U.\subset\al B._u$, and 
for this we will
also use the notation $(\al B.,\,\al U.)$.
\end{defi}
The assumption is that all physical information is contained in the
pair $(\al B.,{\got S}_D)$.
Examples of constraint  theories as defined here, 
have been worked out in detail
for various forms of electromagnetism~cf.~\cite{Grundling85, Grundling88c,
Lledo}.
 
For the case of unitary constraints we have the following equivalent 
characterizations of the Dirac states 
(cf.~\cite[Theorem~2.19~(ii)]{Grundling85}):
\begin{eqnarray}
  \label{DiracU1}{\got S}_{{D}}&=&\Big\{ \omega\in{\got S}({\al B.})\mid 
                   \omega(U)=1 \quad {\forall}\, U\in {\al U.}\Big\} \\[1mm]
  \label{DiracU2}              &=&\Big\{ \omega\in{\got S}({\al B.})\mid 
                   \omega(FU)=\omega(F)=\omega(UF) \quad {\forall}\,
                   F\in\al B.,\; U\in {\al U.}\Big\}.
\end{eqnarray}
Moreover, the set $\{\alpha_U:= {\rm Ad}(U)\mid U\in\al U.\}$ of 
automorphisms of $\al B.$ leaves every Dirac state invariant, 
i.e.~we have $\omega\circ\alpha_U=\omega$ for all 
$\omega\in {\got S}_{{D}}$, $U\in{\al U.}$. 

For a general constraint set $\al C.$, observe that we have:
\begin{eqnarray*}
 {\got S}_D &=& \Big\{ \omega\in{\got S}({\al B.})\mid\omega(C^*C)=0 
                \quad {\forall}\, C\in\al C.\Big\}                \\[1mm]
            &=& \Big\{ \omega\in{\got S}({\al B.})\mid \al C.\subseteq 
           N_\omega\Big\}\kern2mm=\kern2mm\al N.^\perp\cap{\got S}(\al B.)\;.
\end{eqnarray*}
Here $N_\omega:=\{F\in\al B.\mid\omega(F^*F)=0\}$ is the left kernel of
$\omega$ and 
$\al N.:=\cap\; \{N_{\omega}\mid\omega\in{\got S}_D \}$, 
and $\al N.^\perp$ denotes the annihilator of $\al N.$
 in the dual of $\al B.$. 
Now $\al N.=\csp(\al BC.)$
 because every closed 
left ideal is the intersection of the left kernels which contains it 
(cf.~3.13.5 in \cite{bPedersen89}).
Thus $\al N.$ is the left ideal generated by $\al C.$. 
Since $\al C.$ is selfadjoint and contained in 
$\al N.$ we conclude
\[
\al C.\subset {\rm C}^*(\al C.)\subset 
\al N.\cap\al N.^*=\csp(\al BC.)\,\cap\,\csp(\al CB.)\;,
\]
 where ${\rm C}^*(\cdot)$ 
denotes the C*--algebra in $\al B.$ generated by its argument.
Then we have (cf. \cite{Lledo}):

\begin{teo}
\label{Teo.3.1}
For the Dirac states we have:
\begin{itemize}
\item[{\rm (i)}] ${\got S}_{{D}}\neq\emptyset\;$ iff 
   $\;\EINS\not\in {\rm C}^*(\al C.)$ iff 
$\;\EINS\not\in \al N.\cap\al N.^*=:\al D.$.
\item[{\rm (ii)}] $\omega\in {\got S}_D\;$ iff 
   $\; \pi_{\omega}({\al D.})\Omega_{\omega}=0$.
\item[{\rm (iii)}] An extreme Dirac state is pure.
\end{itemize}
\end{teo}

We will call a constraint set $\al C.$ {\bf first class} if
$\EINS\not\in {\rm C}^*(\al C.)$, and this is the nontriviality 
assumption which we henceforth make \cite[Section~3]{Grundling88a}.

Now define 
\[
 {\al O.} := \{ F\in {\al B.}\mid [F,\, D]:= FD-DF \in {\al D.}\quad 
               {\forall}\, D\in{\al D.}\}.
\]
Then ${\al O.}$ is the C$^*$--algebraic analogue of Dirac's observables 
(the weak commutant of the constraints) \cite{bDirac64}.
Then (cf.  \cite{Lledo}): 

\begin{teo}
\label{Teo.2.2}
With the preceding notation we have:
\begin{itemize}
\item[{\rm(i)}] $\al D.=\al N.\cap \al N.^*$ is the unique maximal 
  {\rm C}$^*$--algebra in $\, \cap\; \{ {\rm Ker}\,\omega\mid \omega\in 
  {\got S}_{{D}} \}$. Moreover $\al D.$ is a hereditary 
  {\rm C}$^*$--subalgebra of $\al B.$.
\item[{\rm(ii)}] ${\al O.} = {M}_{\al B.}({\al D.})
  :=\{ F\in{\cal B}\mid FD\in{\cal D}\ni DF\quad\forall\, D\in{\cal D}\}$, 
  i.e.~it is the relative multiplier algebra of ${\al D.}$ in ${\al B.}$.
\item[{\rm(iii)}] $\al O.=\{F\in\al B.\mid\; [F,\,\al C.]\subset\al D.\},$
hence $\al C.'\cap\al B.\subseteq\al O.\;.$
\item[{\rm(iv)}] $\al D.=\csp(\al OC.)=\csp(\al CO.)$.
\item[{\rm(v)}] For the case of unitary constraints, i.e.
 $\al C.=\al U.-\EINS$, we have
 $\al U.\subset\al O.$ and $\al O.={\{F\in\al B.\mid\alpha_U(F)-F\in\al D.
\quad\forall\; U\in\al U.\}}$
  where $\alpha_U:={\rm Ad}\,U$.
\end{itemize}
\end{teo}

Thus $\al D.$ is a closed two-sided ideal of $\al O.$ and it
is proper when ${\got S}_D\not=\emptyset$ (which we assume here
by $\EINS\not\in {\rm C}^*(\al C.)$).
Since the traditional observables are $\al C.'\cap\al B.,$ 
by (iii) we see that these are in $\al O.\,.$ In general 
$\al O.$ can be much larger than $\al C.'\cap\al B..$

Define the {\it maximal {\rm C}$^*$--algebra of physical observables} as
\[
 {\al R.}:={\al O.}/{\al D.}.
\]
The factoring procedure is the actual step of imposing constraints. 
This method of constructing $\al R.$  from $(\al B.,\,\al C.)$
is called the {\bf T--procedure}
in \cite{Grundling85}, and it defines a map $T$ from first class
constraint pairs $(\al B.,\,\al C.)$ to unital C*--algebras by
${T(\al B.,\,\al C.)}:=\al R.=\al O./\al D..$ 
We require that after the T--procedure all physical
information is contained in the pair $({\al R.}\kern.4mm ,{\got S}
({\al R.}))$, where ${\got S}({\al R.})$ denotes the set of 
states on $\al R.$. Now, it is possible that $\al R.$ may not be simple
\cite[Section~2]{Grundling85}, and this would not be acceptable for a 
physical algebra. So, using physical arguments, one would in practice
choose a C$^*$--subalgebra $\al O._c\subset \al O.$ containing 
the traditional observables $\al C.'$ such that
\[
 \al R._c :=\al O._c / (\al D.\cap\al O._c )\subset \al R.\,,
\]
is simple. The following result justifies the choice of $\al R.$ as the
algebra of physical observables (cf. Theorem 2.20 in \cite{Grundling85}):

\begin{teo}
\label{Teo.2.6}
 There exists a ${\sl w}^*\hbox{--continuous}$ isometric bijection 
         between the Dirac states on ${\al O.}$ and the states on ${\al R.}$. 
\end{teo}

Insofar as the physics is
now specified by $\al R.,$ this suggests that we 
call two constraint sets equivalent if they produce the same $\al R..$
More precisely two constraint sets $\al C._1\subset\al B.\supset\al C._2$
are called {\bf equivalent}, denoted $\al C._1\sim\al C._2,$
if they select the same set of Dirac states, cf.~\cite{Lledo}.
In fact
\[
\al C._1\sim\al C._2\quad\hbox{iff}\quad
\csp(\al BC._1)=\csp(\al BC._2)\quad\hbox{iff}\quad\al D._1=\al D._2\;.
\]

The hereditary property of $\al D.$ can be further analyzed, 
and we list the main points (the proofs are
in Appendix~A of \cite{Lledo}).

Denote by $\pi_u$ the universal representation of $\al B.$ on the
universal Hilbert space $\al H._u$ \cite[Section~3.7]{bPedersen89}.
$\al B.''$ is the strong closure of $\pi_u(\al B.)$ and since $\pi_u$
is faithful we make the usual identification of $\al B.$
with a subalgebra of $\al B.''$, i.e.~generally omit explicit
indication of $\pi_u$. If $\omega\in{\got S}(\al B.)$, we will
use the same symbol for the unique extension of $\omega$ from $\al B.$ to 
$\al B.''$. 
\begin{teo}
\label{Teo.2.7}
For a constrained system $(\al B.,\al C.)$ there exists a 
 projection
 $P\in\al B.''$ such that
\begin{itemize}
 \item[{\rm (i)}] $\al N.=\al B.''\,P\cap \al B.$,
 \item[{\rm (ii)}] $\al D.=P\,\al B.''\,P \cap \al B.$ 
 \item[{\rm (iii)}] ${\got S}_D=\{\omega\in{\got S}(\al B.)\mid\omega(P)=0\}$
\item[{\rm(iv)}] $\al O.=\{ A\in\al B. \mid PA(\EINS-P)=0=
                        (\EINS-P)AP \}=P'\cap \al B.\;.$
\end{itemize}
\end{teo}
 A projection satisfying the conditions of Theorem~\ref{Teo.2.7}
is called {\em open} in
\cite{bPedersen89}. 

What this theorem means, is that with respect to
the decomposition 
\[
 \al H._u=P\,\al H._u\oplus (\EINS-P)\,\al H._u
\] 
we may rewrite 
\begin{eqnarray*}
 \al D.&=& \Big\{ F\in\al B.\;\Big|\; F=
          {\left(\kern-1.5mm\begin{array}{cc}
              D \kern-1.6mm & 0 \\ 0 \kern-1.6mm & 0
          \end{array} \kern-1.5mm\right)},\;
       D\in P\al B.P \Big\}\;\;{\rm and}  \\
 \al O.&=& \Big\{ F\in\al B.\;\Big|\; F=
          {\left(\kern-1.5mm\begin{array}{cc}
              A \kern-1.6mm & 0 \\ 0 \kern-1.6mm & B
          \end{array} \kern-1.5mm\right)},\;
      A\in P\al B.P,\; B\in(\EINS-P)\al B.(\EINS-P) \Big\}\,.
\end{eqnarray*}
It is clear that in general $\al O.$ can be much greater than the
traditional observables $\al C.'\cap\al B.$. Next we show how to identify the 
final algebra of physical observables $\al R.$ with a subalgebra of
$\al B.''$.

\begin{teo}
\label{Teo.2.11}
For $P$ as above we have:
\[
 \al R.\cong\Big\{ F\in\al B.\;\Big|\; F=
          \left(\kern-1.5mm\begin{array}{cc}
              0 \kern-1.6mm & 0 \\ 0 \kern-1.6mm & A
          \end{array} \kern-1.5mm\right) \Big\}=
 (\EINS-P)\,(P'\cap\al B.) \subset \al B.''.
\]
\end{teo}

Below we will need to consider a constraint system 
contained in a larger algebra,
specifically, $\al C.\subset\al A.\subset\al F.$ where $\al C.$ is
a first--class constraint set, and $\al A.,\;\al F.$ are unital
C*--algebras.
Now there are 
two constrained systems to consider;-  $(\al A.,\,\al C.)$
and $(\al F.,\,\al C.).$ The first one produces the algebras
$\al D.\subset\al O.\subseteq\al A.,$ and the second produces
$\al D.\s{\al F.}.\subset\al O.\s{\al F.}.\subseteq\al F..$
where as usual, 
\begin{eqnarray*}
\al N.&=& \csp(\al AC.),\qquad \al D.=\al N.\cap\al N.^*,\qquad
\al O.=M\s{\al A.}.(\al D.)
\qquad\hbox{and}  \\
\al N.\s{\al F.}.&=&\csp(\al FC.),\qquad
\al D.\s{\al F.}.=\al N.\s{\al F.}.\cap
\al N.^*\s{\al F.}.,\qquad
\al O.\s{\al F.}.=M\s{\al F.}.(\al D.\s{\al F.}.).
\end{eqnarray*}
Then we have (cf. Theorem~3.2 of \cite{Grundling88b}): 
\begin{teo}
\label{Teo.2.12}
Given as above $\al C.\subset\al A.\subset\al F.$
then
\[ \al N.\s{\al F.}.\cap\al A.=\al N.,\qquad
   \al D.\s{\al F.}.\cap\al A.=\al D.,\qquad
\qquad\hbox{and}\qquad
    \al O.\s{\al F.}.\cap\al A.=\al O.\;.
\]
Hence $\al R.=\al O./\al D.=(\al O.\s{\al F.}.\cap\al A.)\big/
(\al D.\s{\al F.}.\cap\al A.)\;.$
\end{teo}

\section{Superselection with constraints.}

Next we would like to consider systems containing 
both constraints and superselection. 
There is a choice in how to define this problem mathematically,
so let us consider the physical background.
Perhaps the most important
example, is that of a local gauge theory.
It usually has a set of global charges (leading to superselection)
as well as a Gauss law constraint (implementing the local gauge symmetry),
 and possibly also other constraints
associated with the field equation.
Only if the gauge group is abelian will the Gauss law constraint commute with
the global charge, since the Gauss law constraint 
takes its values in the Lie algebra of the gauge group.
Thus, for nonabelian local gauge theories 
we do not expect the constraints to be 
in the algebra of gauge invariant 
observables $\al A.$ of the superselection
theory of the global charge. 
This problem is however not as serious as it looks.
The reason is that whilst the global gauge group does not preserve
the individual Gauss law constraints, it does preserve the set of these,
hence it also preserves the set of Dirac states selected by them.
Thus we can replace the original constraint set by an equivalent constraint
(i.e. selecting the same set
of Dirac states) 
which is invariant under the global gauge group.
Such an equivalent constraint 
is given by the projection in Theorem~\ref{Teo.2.7}.
It comes at the cost of slightly enlarging the system algebra $\al B.,$
since $P$ is in the universal Von Neumann algebra of  $\al B.$.
We can avoid this cost  if $\csp(\al C.)$
is separable, since then there is an equivalent constraint in 
$\al B.$ itself, cf.~Theorem~3.4 of~\cite{Lledo}.

We therefore will assume below that the constraints are in
in $\al A..$ 
This will include the situation where
  there are two or more local gauge symmetries
which mutually commute (e.g. isospin and electromagnetism),
in which case the Gauss law constraint 
of one symmetry will commute with the global charges
of the other.
We can also easily find constraints which are independent of the
gauge symmetries, e.g. restriction to a submanifold,
or enforcing a dynamical law.

Let now $(\al A.,\,\al C.)$ be 
 a first--class constraint system, hence we have the associated
algebras $\al D.\subset\al O.\subseteq\al A.,$ and $\al R.=\al O./\al D..$
In addition, let
$\al A.$ have  a superselection structure
i.e. there is a given Hilbert extension $\HS$ of $\al A..$ 
Thus the category
$\al T.$
of canonical endomorphisms of
$\al A.$
defines a selection criterion of unital endomorphisms of
$\al A..$
In the case that the Hilbert extension is minimal and regular,
the superselection structure of
$\al T.$
is given within
$\al A.$
without any reference to the Hilbert extension.

Then the following natural questions arise:
\begin{itemize}
\item[(1)] what compatibility conditions should be satisfied in order
to pass the superselection structure through $T,$ thus obtaining 
a superselection structure on $T(\al A.,\,\al C.)=\al R.?$
\item[(2)] what is the relation between $T(\al A.,\,\al C.)$
and $T(\al F.,\,\al C.)$ where $\al F.$ is the field algebra generated from
$\al T.?$
\end{itemize}
An inverse question also arises, i.e.
\begin{itemize}
\item[(3)] if $\al R.$ has a superselection structure, what is the
weakest structure one can expect on $\al A.$ which would produce
this superselection structure on $\al R.$ via $T?$
(One should call this a {\it weak} superselection structure.)
\end{itemize}
To address (1) and (2), recall that the map $T$ consists of a restriction
(of $\al A.$ to $\al O.)$ followed by a factoring 
($\al O.\to\al O./\al D.$). So, we first work out the compatibility
conditions involved with restrictions and factoring maps.

Since $\al C.\subset\al A.\subset\al F.\,,$ there are 
two constrained systems to consider;-  $(\al A.,\,\al C.)$
and $(\al F.,\,\al C.).$ The first one produces the algebras
$\al D.\subset\al O.\subseteq\al A.,$ and the second produces
$\al D.\s{\al F.}.\subset\al O.\s{\al F.}.\subseteq\al F.$
(cf. Theorem~\ref{Teo.2.12}).
Now since $\al C.\subset\al A.,$ the $\al G.\hbox{--invariant}$ part
of $\al F.,$ it follows that $\al G.$ preserves the set of Dirac states,
hence $\al G.$
 preserves both
 $\al D._{\al F.}$
 and
 $\al O._{\al F.}$,
 i.e.
 $g\al D._{\al F.}=\al D._{\al F.}$
 and
 $g\al O._{\al F.}=\al O._{\al F.}$
 for all
 $g\in\al G.$.

We denote the restriction of
$\al G.$
to
$\al O._{\al F.}$
by
$\beta_{g}:={g}\rest\al O._{\al F.}.$
The homomorphism
$\beta:\al G.\ni g\rightarrow \beta_{g}\in\aut\,\al
O._{\al F.}$
is not necessarily injective but
$\beta$
is again
pointwise norm-continuous, hence
$\al G./\al K.$
is compact where
$\al K.:=\ker\beta.$
The isomorphism
$\tilde{\beta}:\al G./\al K.\to
\beta_{\al G.}$ by
$\tilde{\beta}(g\al K.):=\beta_{g}$
is also a topological one (cf.~p.58~\cite{HR}).
Note that 
$\wh{(\al G./\al K.)}=\{\gamma\in\wh{\al G.}\,\bigm|\,\gamma(k)=1\quad 
\hbox{for all}\;k\in\al K.\}\supseteq\spec\,\beta_{\al G.}\,.$

The spectral projections 
$\Pi^{\beta}_{\gamma}$
of
$\beta\s{\al G.}.$
are given by the restriction to
$\al O._{\al F.}$
of the spectral projections 
$\Pi_{\gamma}$ 
of
$\al G.$,
i.e. 
$\Pi^{\beta}_{\gamma}X=\Pi_{\gamma}X$
for
$X\in\al O._{\al F.}.$
 
We now have the:
\begin{itemize}
\item[(I)] 
Restriction problem. 
Find conditions to 
guarantee that 
the C*--dynamical system
$\{\al O._{\al F.},\al G.,\beta\}$
is a Hilbert system 
$\{\al O._{\al F.},\beta_{\al G.}\}$.
Thus we have to find conditions to ensure there are algebraic
Hilbert spaces in
$\Pi^{\beta}_{\gamma}\al O._{\al F.}$
for
$\gamma\in\wh{(\al G./\al K.)}$.
(Note that this is stronger than what we need;-
we only need a Hilbert system on $\al R.\s{\al F.}.$ after factoring 
out by $\al D.\s{\al F.}..)$ 
\item[(II)] 
Factoring problem. 
Find conditions to guarantee that under the map
 $\al O.\s{\al F.}.\to\al R.\s{\al F.}.:={\al O.\s{\al F.}.\big/
\al D.\s{\al F.}.}$ the factoring through  of the action of $\al G.$ 
to $\al R.\s{\al F.}.$
is a Hilbert system corresponding to a DR--category. 
This is of course a special case of the general
problem for homomorphic images of Hilbert systems
under factoring by invariant ideals.
The reason why we require $Z(\al A.)=\C\un$ for $\al R.\s{\al F.}.$
is because after implementing constraints, the final physical algebra
should be simple.
\end{itemize}
Below we list our major results;- since some proofs are lengthy, 
we defer these to  Section~\ref{Proofs} to preserve the main flow of ideas.

\subsection{Restricting a superselection structure.}
\label{restrict}

We consider now for the system above the restriction problem (I), 
i.e. we are given a Hilbert extension $\HS$ of $\al A.,$ containing
constraints $\al C.\subset\al A.,$ and we need to examine
when 
$\{\al O._{\al F.},\,\beta\s{\al G.}.\}$
is a Hilbert system.
\def\of{\al O.\s{\al F.}.}
\def\df{\al D.\s{\al F.}.}
\def\rf{\al R.\s{\al F.}.}
\begin{teo}
\label{Teo.4.1}
\begin{itemize}
\item[(i)] $\{\al O._{\al F.},\al G.,\beta\}$
has fixed point algebra $\al O..$ Moreover, $Z(\al A.)\subseteq
Z(\al O.).$ 
\item[(ii)] 
For any $\al G.$-invariant
algebraic Hilbert space  $\al H._\gamma\subset\Pi_\gamma\al F.$
we have either
$\al H._\gamma\cap\of=\{0\},$ 
or $\al H._\gamma\subset\of\,.$
In the latter case we have 
$\gamma\in 
\wh{\al G./\al K.}$
where $\al K.=\ker\beta\,,$
and
\[
\al H._\gamma\subset\Pi_{\gamma}\of=\csp(\al O.\al H._{\gamma}).
\]
\item[(iii)]
Let $\sigma\in\ob\al T.,$ with $\al H._\sigma\subset\al F.$
a $\al G.$-invariant algebraic Hilbert 
space such that
$\sigma=\rho\s{\al H._\sigma}..$ 
If $\al H._\sigma\subset\of,$
then $\sigma(\al D.)\subseteq\al D.$ and $\sigma(\al O.)\subseteq\al O..$ 
Thus $\sigma$ restricts to $\al O.,$ $\sigma\rest\al O.\in\endo
\al O..$
\end{itemize}
\end{teo}

The central condition for
$\{\of,\al G.,\beta\}$
to be a Hilbert system
$\{\of,\beta_{\al G.}\}$
w.r.t. the factor group
$\al G./\al K.$
is
$\al H._{\gamma}\subset\of$,
i.e.
$\al H._{\gamma}\subset\Pi_{\gamma}^{\beta}\of$
for all
$\gamma\in\wh{\al G./\al K.}$.

Next, we develop an internal criterion on $\al A.$ to guarantee that
a given
$\al H.\in\ob\al T._{\al G.}$
is contained in
$\of$.
 
\begin{teo}
\label{Teo.4.2}
\begin{itemize}
\item[(i)]
Given the Hilbert extension $\HS$ of the constrained system
$\al C.\subset\al A.$ assumed here, we have for
any $\al G.\hbox{--invariant}$ algebraic Hilbert space $\al H.$
 that 
\begin{eqnarray*}
\al H.\subset\of\qquad &\hbox{iff}&\qquad
\al D.\sim\rho\s{\al H.}.(\al D.)\\[1mm]
&\hbox{i.e.}&\qquad
\al D.=\csp\big(\al A.\rho\s{\al H.}.(\al D.)\big)\,\cap
\,\csp\big(\rho\s{\al H.}.(\al D.)\al A.\big)\,.
\end{eqnarray*}
\item[(ii)]
For all $\sigma,\;\tau\in\ob\al T.$
 with $\al H._\sigma,\al H._\tau\subset\of$ we have
 \[
 (\sigma,\,\tau)\s{\al A.}.\subseteq\big(\sigma\rest\al O.,\,\tau
 \rest\al O.\big)\s{\al O.}.\;.
 \]
\end{itemize}
\end{teo}
Observe  
that $\al D.\sim\rho\s{\al H.}.(\al D.)$ implies that 
$\rho\s{\al H.}.(\al D.)\subseteq\al D..$
\begin{cor}
\label{cor4.1}
We have that $\{\of,\al G./\al K.,\tilde{\beta}\}$ is a Hilbert
system
$\{\of,\beta_{\al G.}\}$
w.r.t.
$\al G./\al K.$ iff
$\al D.\sim\rho_\gamma(\al D.)$ 
holds for all 
$\gamma\in\wh{\al G./\al K.}.$ 
In particular, if
$\al D.\sim\rho_{\gamma}(\al D.)$
holds for all
$\gamma\in\wh{\al G.}$
then
$\al G./\al K.\cong\al G.$
i.e. $\al K.$ is trivial.
\end{cor}

Whilst the condition $\al D.\sim\rho\s\gamma.(\al D.)$  
 is exact for
$\al H._\gamma\subset\of,$ it may not be in practice
that easy to verify.
We therefore consider alternative conditions which will
allow the main structures involved with Hilbert extensions
to survive the restriction of
$\HS$ to $\{\of,\,\beta\s{\al G.}.\}.$
 
Recalling the definition of subobjects, introduce the notation
$E\simeq \un({\rm mod}\,\al A.)$
for a projection $E\in\al A.$ to mean that
there is an isometry $V\in\al A.,$ $V^*V=\un$ such that
$VV^*=E$ (i.e. Murray--Von Neumann equivalence of $E$ and $\un).$

\begin{defi}
We say the constraint set $\al C.\subset\al A.$ is an 
{\bf E--constraint set} if for each projection $E\in\al O.$
such that $E\simeq \un({\rm mod}\,\al A.),$ we have 
that $E\simeq \un({\rm mod}\,\al O.).$
\end{defi}

The E-constraint condition will ensure the
survival of decomposition relations of restrictable canonical 
endomorphisms:

\begin{pro}
\label{pro.4.7}
Let 
$\HS$ be a Hilbert system and let
$\al C.\subset\al O.$ be an E--constraint set, 
$\sigma\in\ob\al T.$
and
$\al H._\sigma\subset\of$   
a 
$\al G.$-invariant
algebraic Hilbert space. Then
\begin{itemize}
\item[(i)] to each decomposition
\[
\sigma(\cdot)=\sum_jV_j\rho\s\gamma_j.(\cdot)
V_j^*\;,\qquad V_j\in(\rho\s\gamma_j.,\,\sigma)\s{
\al A.}.\;,
\]
where $\gamma_j\in\wh{\al G.}$ and $V_j\in\al A.$ are isometries, 
there corresponds a decomposition on
$\al O.,$ i.e.
there are 
$\al G.$-invariant
algebraic Hilbert spaces
$\al K._j\subset\of,$ 
which carry the representation $\gamma_j$ and
with  canonical endomorphisms $\tau_j
:=\rho_{\al K._{j}}\rest\al O.\in\endo\al O.$ 
such that
on $\al O.:$
\[
\sigma(\cdot)=\sum_jW_j\tau_j(\cdot)
W_j^*\;,\qquad W_j\in(\tau_j,\,\sigma)\s{
\al O.}.\;,
\]
where $W_j\in\al O.$ are isometries.
\item[(ii)] 
Let 
$\{\al F.,\al G.\}$
in addition satisfies Property B and let
$\tau<\sigma\in\al T.$ in the sense of $\al A.,$
i.e. there is an isometry $V\in{(\tau,\,\sigma)\s{\al A.}.},$
and let $\al H._\sigma\subset\of.$
Then 
there is a corresponding Hilbert space
$\al H._\tau\subset \of$ i.e.
$\tau\rest\al O.<\sigma\rest\al O.\in\endo\al O.$
also in the sense of $\al O..$
\end{itemize}
\end{pro}

\begin{teo}
\label{tensrep}
Let 
the Hilbert system
$\HS$ 
satisfy Property B
where $\al G.$ is a group 
with a distinguished  
irreducible representation $\gamma_0\in\wh{\al G.}$
such that every irreducible representation of $\al G.$  
is contained in a tensor representation of $\gamma_0.$
Let $\al C.\subset\al A.$ be an E--constraint set
then
$\al H.\s\gamma_0.\subset\of$ implies that 
$\{\of,\beta\s{\al G.}.\}$ is a Hilbert system.
\end {teo}
\begin{beweis}
This follows from Proposition~\ref{pro.4.7}, by making use of the 
obvious fact that $\al H._\tau\subset\of\supset\al H._\sigma$
implies  that $\al H._\tau\cdot\al H._\sigma\subset\of$
for $\sigma,\;\tau\in\ob\al T.\,.$
\end{beweis}
If the group 
$\al G.$
is isomorphic to $U(N)$ then it
satisfies the condition of Theorem~\ref{tensrep}. 

The property of being  an E--constraint set can be 
characterized in terms of the open projection 
$P\in\al A.''$  corresponding to the
constraints (cf. Theorem~\ref{Teo.2.7}). 
Observe that if there is an $E\in\al O.$
with $E\simeq\un({\rm mod}\,\al A.),$
then the set of isometries
\[
\al V._E:=\set V\in\al A., VV^*=E,\quad V^*V=\un.
\]
is nonempty. We have:
\begin{pro}
Let $E\in\al O.$
with $E\simeq\un({\rm mod}\,\al A.),$ then $\al V._E\cap\al O.\not=
\emptyset$ iff for each $V\in\al V._E$ there is a
\[
 U\in \al U._E:=\set U\in\al A., U^*U=E=UU^*.
\]
such that $VPV^*=UPU^*.$
\end{pro}
  
\subsection{Morphisms of general Hilbert systems.}
\label{morphsec}

Recall that the second step in the enforcement of constraints, is
the factoring $\of\to\rf:=\of/\df.$ We now consider 
problem {(II)}, the factoring problem, first in a general
context. Consider a  morphism of 
C*--algebras  $\xi:\al F.\to\al L.=\xi(\al F.).$
This specifies the subgroup of
automorphisms
\[
\autx\al F.:=\set\alpha\in\aut\al F.,\alpha(\ker\xi)\subseteq
\ker\xi.
\]
and a homomorphism $\autx\al F.\to\aut\al L.$ by
$\alpha\to\alpha^\xi$ where $\alpha^\xi(\xi(F)):=
\xi(\alpha(F))$ for all $F\in\al F..$
Henceforth let $\HS$ be a Hilbert system with Property B
and ${\al G.}\subset\autx\al F..$
Our task will be to find the best conditions to
ensure that $\{\al L.,\,\al G.^{\xi}\}$
is a Hilbert system associated with a category
described in Theorem~\ref{Teo1}.
We will denote the spectral projections
of $\al G.$ (resp. $\al G.^\xi)$ by  $\Pi_\gamma$
(resp. $\Pi^\xi_\gamma).$ 
(Recall that in the context of the T--procedure, we have that
$\al G.$ preserves $\df$ due to the invariance of the constraints
under $\al G..$ So the current analysis applies).

\begin{teo}
\label{Teo.4.3}
Given a Hilbert system $\HS$ 
and a unital
morphism $\xi:\al F.\to\al L.=\xi(\al F.),$
such that ${\al G.}\subset\autx\al F.,$ then
we have:
\begin{itemize}
\item[(i)] $\{\al L.,\,\al G.^\xi\}$ is a Hilbert system
and $\al G.\cong\al G.^{\xi}.$
\item[(ii)] If $\al H._\gamma\subset\Pi_\gamma\al F.$ 
is an invariant algebraic Hilbert space 
for $\al G.,$
then so is $\xi(\al H._\gamma)\subset\Pi^\xi_\gamma\al L.$
for  $\al G.^\xi.$
\item[(iii)] Let $\al N._\gamma$ be any
orthonormal basis for $\xi(\al H._\gamma),$
then $\bigcup\set\al N._\gamma,\gamma\in\wh{\al G.}.$ is a
left module basis of $\xi(\al F._{\rm fin})$ w.r.t.
$\xi(\al A.),$ i.e. the ``essential part'' of $\xi$
is its action on $\al A..$
\item[(iv)] The fixed point algebra of $\al L.$
w.r.t.
$\al G.^{\xi}$
 is exactly $\xi(\al A.),$ and
$\xi(\al F._{\rm fin})=\al L._{\rm fin}\;.$
\item[(v)] If $\HS$ has Property B, so does
 $\{\al L.,\,\al G.^\xi\}.$
\end{itemize}
\end{teo}
Thus corresponding to the two Hilbert systems
$\HS$ and $\{\al L.,\al G.^{\xi}\}$
we now have the two categories $\al T.$ and
$\al T.^\xi$ respectively. Moreover:
\begin{cor}
\label{Cor.4.9}
Under the conditions of Theorem~\ref{Teo.4.3} we have that
\begin{itemize}
\item[(i)]
for any canonical endomorphism $\lambda\in\ob\al T.,$
\[
 \lambda(\ker\xi\cap\al A.)\subseteq\ker\xi\cap\al A.\;.
\]
Hence there is a well--defined map
$\ob\al T.\ni\lambda\to\lambda^\xi\in\ob\al T.^\xi,$
given by $\lambda^\xi(\xi(A)):=\xi(\lambda(A))$ for all 
$A\in\al A..$
\item[(ii)] the map $\ob\al T.\ni\lambda\to\lambda^\xi\in\ob\al T.^\xi$
is compatible with products, direct sums and subobjects.
It also preserves unitary equivalence.
\end{itemize}
\end{cor}
We have that $\big(\ob\al T.\big)^\xi\subseteq\ob\al T.^\xi,$
and we now claim that up to unitary equivalence, we have in fact
equality:
\begin{teo}
\label{Teo.4.4} 
Under the conditions of Theorem~\ref{Teo.4.3} we have that
\begin{itemize}
\item[(i)]
if $\sigma\in\ob\al T.^\xi,$ then there is always a
$\lambda\in\ob\al T.$ such that $\lambda^\xi$ is unitarily
equivalent to $\sigma,$ i.e. each unitary equivalence class in
$\ob\al T.^\xi$ contains at least one element of the form 
$\lambda^\xi.$
\item[(ii)]
the map $\ob\al T.\ni\lambda\to\lambda^\xi\in\ob\al T.^\xi$
produces an isomorphism between the sets of unitary equivalence
classes of $\ob\al T.$ and $\ob\al T.^\xi$ which
is compatible with products direct sums and subobjects.
\end{itemize}
\end{teo}
The relation between the arrows of the two categories
is however less direct:
\begin{lem}
\label{arrow1}
Under the conditions of Theorem~\ref{Teo.4.3} we have
\[
\xi\big((\sigma,\,\tau)_{\al A.}\big)\subseteq
\big(\sigma^\xi,\,\tau^\xi\big)\s{\xi(\al A.)}.\;.
\]
\end{lem}
Next we show that $\ker\xi$ is uniquely
determined by $\ker\xi\cap\al F._{\rm fin}.$
\begin{pro} 
\label{Prop.4.12}
Under the conditions of Theorem~\ref{Teo.4.3} we have that
\begin{itemize}
\item[(i)]
$\ker\xi\cap\al F._{\rm fin}=
{\rm Span}\set(\ker\xi\cap\al A.)\al H._\gamma,
\gamma\in\wh{\al G.}.\;,$
\item[(ii)]
$\ker\xi=\clo\s{\vert\cdot\vert_{\al A.}}.
(\ker\xi\cap\al F._{\rm fin})
\cap\al F.\;.$
\end{itemize}
\end{pro}
Thus $\ker\xi$ is in fact uniquely determined by $\ker\xi\cap\al A.,$
as is already suggested by Theorem~\ref{Teo.4.3}(iii).
Since $\al F.$ is in general not complete w.r.t. $|\cdot|_{\al A.},$
the intersection with $\al F.$ in \ref{Prop.4.12}(ii) is necessary.

Theorem~\ref{Teo.4.3} suggests that we consider the following
subcategory 
of
$\al T.^{\xi}$.

\begin{defi}
\label{CatXT}
The subcategory
$\xi(\al T.)$
of
$\al T.^{\xi}$
is defined by the objects
\[
\ob\xi(\al T.):=(\ob\al T.)^{\xi}
\]
and the arrows
\[
(\sigma^{\xi},\tau^{\xi})_{\xi(\al A.)}^{(0)}:=
\xi((\sigma,\tau)_{\al A.}).
\]
\end{defi}

By Theorem~\ref{Teo.4.4} the sets of all unitary
equivalence classes of
$\hbox{Ob}\,\xi(\al T.)$
and
$\hbox{Ob}\,\al T.^{\xi}$
coincide,  each equivalence class of
$\hbox{Ob}\,\xi(\al T.)$
is a subset of the corresponding equivalence class of
$\hbox{Ob}\,\al T.^{\xi}$,
but in general these equivalence classes are much larger.

Lemma~\ref{arrow1} says that the arrow sets
$(\sigma^{\xi},\tau^{\xi})_{\xi(\al A.)}$
of the objects of
$\hbox{Ob}\,\xi(\al T.)$
considered as objects of
$\hbox{Ob}\,\al T.^{\xi}$
are in general larger than the corresponding arrow sets in
$\xi(\al T.)$.
The reason is that an element
$X=\xi(Y),\,Y\in\al A.,$
belongs to
$(\sigma^{\xi},\tau^{\xi})_{\xi(\al A.)}$
iff
$Y\sigma(A)-\tau(A)Y\in\ker\,\xi$
for all
$A\in\al A.$.
The arrow sets coincide only if this relation already implies
$Y\sigma(A)-\tau(A)Y=0.$

\subsection{Morphisms of minimal and regular Hilbert systems }

Recall now that by Theorems \ref{Teo1} and \ref{Teo2} we have an
equivalence between 
minimal and regular Hilbert systems  
with Property B
and the endomorphism
category $\al T.$ with an admissible  
subcategory $\al T._\C.$  
We called a subcategory $\al T._\C$
{\it admissible} if it satisfies conditions P.1--P.2 in Theorem~\ref{Teo1}.

As in the last subsection, we consider a unital morphism
$\xi:\al F.\to\al L.=\xi(\al F.),$ and recall by Proposition~\ref{Prop.4.12}
that $\xi$ is determined by its action on $\al A..$
Now whilst it is obvious that $\xi(Z(\al A.))\subseteq Z(\xi(\al A.)),$
we require below the stronger condition:
\begin{equation}
\label{ZZ}
\xi(Z(\al A.))=Z(\xi(\al A.))\;.
\end{equation}
When $\xi(\al A.)$ is a simple C*--algebra (as we require for
the final observables after a T--procedure), the condition~(\ref{ZZ})
will be satisfied.
\begin{teo}
\label{Teo.4.13}
Given a minimal and regular Hilbert system $\HS$ 
with Property B,
and a unital morphism $\xi:\al F.\to\al L.=\xi(\al F.)$
such that ${\al G.}\subset\autx\al F.$ and condition~(\ref{ZZ})
holds, then:
\begin{itemize}
\item[(i)] there is a DR- subcategory
$\al T.^\xi_\C$ of $\xi(\al T.)$,
\item[(ii)] property P.2 is satisfied for $\al T.^\xi_\C$ 
iff
$\xi(\al A.)'\cap\xi(\al F.)=\xi(Z(\al A.)).$
In this case the subcategory $\al T.^\xi_\C$
is admissible.
\item[(iii)] If $\xi(\al A.)'\cap\xi(\al F.)=\xi(Z(\al A.)),$
 then 
\[
\big(\sigma^\xi,\,\tau^\xi\big)\s{\xi(\al A.)}.
= \xi\big((\sigma,\,\tau)\s{\al A.}.\big)
\]
for all $\sigma,\;\tau\in\ob\al T.,$ where we made use of the notation 
and result in Corollary~\ref{Cor.4.9}.

\item[(iv)] In this case choose
$\al H._{\gamma}\in\ob\al T._{\al G.}\,.$
 Then
\[
\al M.^{\xi}:=\set{\rho\s{\xi(\al H._{\gamma})}.},{\gamma\in\wh{\al G.}}.
\subset\ob\xi(\al T.)
\]
is a complete system of (irreducible) and mutually disjoint
objects of
$\ob\xi(\al T.).$
\end{itemize}
\end{teo}

\subsection{The inverse problem.}

\begin{teo}
\label{Teo.4.14}

Let $\al A.$ be a unital C*--algebra with Property B, and let $\al T.$ be 
a C*-tensor
category  of 
unital endomorphisms of
$\al A.$ .
Let
$\al T.$ have an  
admissible subcategory
$\al T._{\C}$
whose arrow spaces are denoted by
$(\sigma,\tau)_{\C}$.
Furthermore, let $\xi$
be a unital morphism of $\al A.$ such that
\begin{itemize}
\item[(i)] 
$\xi(Z(\al A.))=Z(\xi(\al A.)),$
\item[(ii)] 
$\lambda(\ker\xi)\subseteq\ker\xi$  for all $\lambda\in\ob\al T..$
Thus we can define endomorphisms $\lambda^\xi\in\endo\xi(\al A.)$ by
$\lambda^\xi\big(\xi(A)\big):=\xi\big(\lambda(A)\big)$ for all $A\in\al A.$
and a category
$\xi(\al T.)$
with objects
\begin{equation}
\label{Obxi}
\ob\,\xi(\al T.):=\set{\lambda^\xi},{\lambda\in\ob\,\al T.}.
\end{equation}
and arrows
$(\sigma^{\xi},\tau^{\xi})_{\xi(\al A.)}$,
which is closed w.r.t. direct sums and products.
\item[(iii)] 
$\xi\big((\sigma,\,\tau)_{\al A.}\big)=
\big(\sigma^\xi,\,\tau^\xi\big)\s{\xi(\al A.)}.$ for all
$\sigma,\;\tau\in\ob\al T..$
\end{itemize}

Then there is a subcategory
$\al T.^{\xi}_{\C}$
of
$\xi(\al T.)$
with
$\ob\al T.^{\xi}_{\C}=\ob\xi(\al T.)$
which is admissible for
$\xi(\al T.).$

Thus by Theorem \ref{Teo2} there are Hilbert 
extensions $\al F.$ and $\al F.^\xi$
corresponding to $\al T.$ and $\xi(\al T.)$ repectively.
Moreover, the Hilbert extension $\al F.^\xi$ of $\xi(\al A.)$ can be chosen 
in such a way that it is the homomorphic image of $\al F.$ under
a morphism which is an extension of $\xi.$ 
That is, $\al F.^\xi=\wt\xi(\al F.)$ where $\wt\xi$ is a morphism
of $\al F.$ such that $\wt\xi(A)=\xi(A)$ for all $A\in\al A..$
\end{teo}
\begin{rem}
A posteriori, the set of objects $\ob\xi(\al T.)$ defined in (\ref{Obxi})
could be enlarged by filling up the unitary equivalence classes of
each $\lambda^\xi$ by {\it all} $\tau$ with
$\tau=\Ad V\circ\xi(\lambda),$ where $V\in\xi(\al A.)$ is unitary.
This corresponds to the objects of the category
$\al T.^\xi$
of Definition~\ref{CatXT}
In this case we also have to add additional arrows, so if
$\tau_i=\Ad V_i\circ\xi(\lambda_i),$ $i=1,2,$ then we also need
\[
  (\tau_1,\,\tau_2)\s{\xi(\al A.)}.:=
V_2\big(\tau_1^\xi,\,\tau_2^\xi\big)\s{\xi(\al A.)}.V_1^{-1}\;.
\] 
However, for the application of Theorem~\ref{Teo2} this is not 
necessary.
\end{rem}

\subsection{Superselection structures left after constraining.}

Recall that the enforcement of constraints by T--procedure produces a
final physical algebra $\al R..$ This algebra is usually assumed to be simple;-
if it is not, then the physics is not fully defined, 
and one should extend the constraint
set $\al C.\subset\al A.$ to make $\al R.$ simple (the choice of the extension
needs to be physically motivated).

In the previous subsections we examined which conditions need to be satisfied 
by a Hilbert extension ${\{\al F.,\,\al G.\}}$ of $\al A.$ for its structure
to pass through the two parts of the T--procedure.
Here we combine these to produce conditions on the initial system which will
ensure that we obtain a Hilbert extension of $\al R..$
We will also examine when this final Hilbert extension is regular
(and this produces then a DR--category via simplicity of $\al R.).$

\begin{teo}
\label{HExR}
Let  ${\{\al F.,\,\al G.\}}$ be a Hilbert extension of $\al A.,$ and let
$\al C.\subset\al A.$ be a first-class constraint set such that
$\al D.\sim\rho_\gamma(\al D.)$ 
holds for all 
$\gamma\in\wh{\al G./\al K.}.$
Then ${\{\al R.\s{\al F.}.,
\,\beta\s{\al G.}.\}}$
is a Hilbert extension of $\al R.,$ where $\al R.\s{\al F.}. 
=\xi(\al O.\s{\al F.}.),$
and $\xi$ is the factor map $\al O.\s{\al F.}.
\to \al O.\s{\al F.}.\big/\al D.\s{\al F.}..$
\end{teo}
\begin{beweis}
By Corollary~\ref{cor4.1} it follows from the hypotheses that 
${\{\al O.\s{\al F.}.,\,\beta\s{\al G.}.\}}$
is a Hilbert extension of $\al O..$
Since the constraint set $\al C.\subset\al A.$ is $\al G.\hbox{--invariant,}$
we have that $\alpha(\al D.\s{\al F.}.)=\al D.\s{\al F.}.$
for all $\alpha\in\beta\s{\al G.}.\subset\aut\al O.\s{\al F.}.,$
i.e. $\beta\s{\al G.}.\subset\autx\al O.\s{\al F.}..$ (Recall the discussion
in the introductory part of Section~4).
Thus by Theorem~\ref{Teo.4.3} it follows that
$(\beta\s{\al G.}.)^\xi\cong\beta\s{\al G.}.,$ and that
\[
\left\{\xi\big(\al O.\s{\al F.}.),\,(\beta\s{\al G.}.)^\xi\right\}
=\left\{\al R.\s{\al F.}.,\,\beta\s{\al G.}.\right\}
\]
is a Hilbert extension of $\al R.=\xi(\al O.).$
\end{beweis}

Next, we would like to examine when a 
Hilbert extension as in Theorem~\ref{HExR}
will produce a 
minimal and regular Hilbert extension of $\al R.$
(with Property B).

First recall the requirement for a Hilbert system
$\{\al F.,\,\al G.\}$ to be regular:
there is an 
assignment $\sigma\to\al H._\sigma$ from $\ob\al T.$ to 
$\al G.$-invariant algebraic Hilbert spaces 
in $\al F.$ such that
\begin{itemize}
\item[(i)] $\sigma=\rho\s{\al H._\sigma}.,$ 
i.e. $\sigma$ is the canonical endomorphism of $\al H._\sigma,$
\item[(ii)] $\sigma\circ\tau\to\al H._\sigma\cdot\al H._\tau\;,$
\end{itemize}
that is, the assignment is compatible with products.

We now want to check whether this property also survives the
map $T:\{\al F.,\,\al G.\}\longrightarrow
{\{\al R.\s{\al F.}.,\,\beta\s{\al G.}.\}}. $

\begin{pro}
\label{Tsat}
Let $\al T.$ 
satisfy regularity. Let
 $\al D.\sim\rho_\gamma(\al D.)$ 
for all 
$\gamma\in\wh{\al G./\al K.},$ then
${\{\al R.\s{\al F.}.,\,\beta\s{\al G.}.\}}$
satisfies regularity, i.e.
there is an 
assignment $\sigma\to\al H._\sigma$   
 such that
\begin{itemize}
\item[(i)] $\sigma=\rho\s{\al H._\sigma}.,$ 
i.e. $\sigma$ is the canonical endomorphism of $\al H._\sigma,$
\item[(ii)] $\sigma\circ\tau\to\al H._\sigma\cdot\al H._\tau\;.$
\end{itemize}
\end{pro}
\begin{beweis}
Given the assignment $\sigma\to\al H._\sigma$ in $\al F.,$
then whenever $\sigma=\rho_{\gamma},$ $\gamma\in\wh{\beta\s{\al G.}.}$
we have 
\[
   \sigma\to\al H._\sigma\subset\al O.\s{\al F.}.\longrightarrow
   \al R.\s{\al F.}.
\]
where the last map is $\xi,$ so the assignment which we take for
this proposition is $\sigma\to\xi(\al H._\sigma).$
Then (i) and (ii) are automatic. 
\end{beweis}

Second, we consider Property B.

\begin{pro}
Let 
$\{\al F.,\al G.\}$ satisfy Property B, let $\al G.$ be nonabelian  and
$\al C.\subset\al A.$
be an E-constraint set. If
$\al D.\sim\rho_{\gamma}(\al D.)$
for all
$\gamma\in \wh{(\al G./\al K.)},$
then
$\{\al O._{\al F.},\beta_{\al G.}\}$
satisfies Property B.
\end{pro}

\begin{beweis}
First
$\{\al O._{\al F.},\beta_{\al G.}\}$
is a Hilbert extension of
$\al O.$ 
w.r.t.
$\al G./\al K.$
because of Corollary~\ref{cor4.1}. Choose an
$\al G.\hbox{-invariant}$ Hilbert space
$\al H.\subset\al O._{\al F.}\subset\al F.$
which is not irreducible, i.e. there is
a projection
$E\in\al J.(\al L._{\al G.}(\al H.)),\,0<E<\un.$
Then one has
$E\in (\rho_{\al H.},\rho_{\al H.})_{\al A.}\subset
(\rho_{\al H.}\rest\al O.,\rho_{\al H.}\rest\al O.)_{\al O.}\subset\al O.$
by Theorem~\ref{Teo.4.2}(ii).
By Property B we get closure under subobjects, so there is a
$V\in\al A.,\,V^{\ast}V=\un,\,VV^{\ast}=E.$
In other words,
$E\cong\un(\hbox{mod}\,\al A.).$
Similarly we obtain
$\un-E\cong\un(\hbox{mod}\,\al A.).$
Since
$\al C.$
is an E-constraint set and $E\in\al O.$ we get that
$E\cong\un(\hbox{mod}\,\al O.)$
and
$\un-E\cong\un(\hbox{mod}\,\al O.)$
and this is the assertion.
\end{beweis}
 
Finally, we need to consider whether the requirement
\[
  \al A.'\cap\al F.=Z(\al A.)
\]
passes through the T-procedure.
In full generality, this is a very hard
problem, because both stages of the T-procedure can eliminate or create
elements of $\al A.'.$ 
In fact, since $\al A.'\cap\al F.\subset\al D.'\cap\al F.\subset
\al O.\s{\al F.}.$ and $Z(\al A.)\subset Z(\al O.),$
we can only deduce from  $\al A.'\cap\al F.=Z(\al A.)$
that $\xi(\al A.'\cap\al F.)=\xi(Z(\al A.)).$
On the other hand,
  $\al R.'\cap\al R.\s{\al F.}.=Z(\al R.)$ iff
\[
  A\in\al O.\s{\al F.}.\quad\hbox{and}\quad [A,\,\al O.]
 \subset\al D.\s{\al F.}.\quad\hbox{implies}\quad
 A\in \al O.+\al D.\s{\al F.}.\,
\]
which can be true in general for more
elements than those in $\xi(\al A.'\cap\al F.).$

We do have from Theorem~\ref{Teo.4.2}  and Proposition~\ref{disjZ}
the following condition:

\begin{pro}
Let
$\HS$
be a minimal Hilbert extension of
$\al A.,$ and let $\al C.\subset\al A.$ be
a first-class constraint set such that
$\al D.\sim\rho_\gamma(\al D.)$ 
holds for all 
$\gamma\in\wh{\al G./\al K.}.$
If the disjointness of
canonical endomorphisms survives the restriction to
$\al O.$
then the Hilbert system
$\{\of,\beta_{\al G.}\}$
is minimal, i.e.
$\al O.'\cap\of=\al Z.(\al O.)$.
\end{pro}

\section{Example}

It is difficult to produce interesting worked examples 
in the current state of the theory. The problem is that in almost all
theories of physical significance, the canonical endomorphisms 
$\rho_\gamma$ are not known explicitly, and so one cannot
check the compatibility conditions with the constraints 
explicitly (cf. Corollary~\ref{cor4.1}). 
Here we give an example which is extracted from QED, so it may have
some physical interest. It consists of a fermion in an Abelian gauge
potential. Since the global gauge group  $\al G.$
is abelian, the superselection theory simplifies radically.
However, we have explicit endomorphisms $\rho_\gamma$ and can check the
compatibility conditions with the constraints.
Nevertheless, even at this simple level, it is not possible to verify all the
 conditions of regularity. We will not treat the issue of dynamics.

\subsection{Constraint structure of QED}

We start with a discussion of the set--up of QED 
in order to motivate our subsequent example.
The starting point for QED, is a fermion field $\psi$ in $\R^4$ satisfying
the free CARs, and a $U(1)\hbox{--gauge}$ potential $A$ in $\R^4$ satisfying 
free CCRs, and initially these are assumed to commute. So the
appropriate C*-algebraic framework at this initial level is
\[
\al B.:={\rm CAR}(\al H.)\otimes{\rm CCR}(S,\,B)
\]
where $\al H.=L^2(\R^4,\,\C^4),$ $S=\al S.(\R^4,\,\R^4)\big/\ker B,$
and $B$ denotes the symplectic form for QEM, coming from the
Jordan--Wigner distribution, cf. Sect~5 of \cite{Lledo}.
 (Note that the tensor product $\al B.$
is unique because ${\rm CAR}(\al H.)$ is a nuclear algebra.)
There is a global charge $Q$ acting on ${\rm CAR}(\al H.)$
and there are constraints in the heuristic theory:
\begin{eqnarray*}
{A_\mu}^{,\,\mu}(x)&:=&0  \quad\qquad\hbox{(Lorentz condition)} \\[1mm]
\hbox{and}\quad\qquad
\Box A_\mu&:=&j_\mu \quad\qquad\hbox{(Maxwell equation)}
\end{eqnarray*}
where $j_\mu:=-e\wt\psi\gamma_\mu\psi$ is the electron current,
and we denote $\wt\psi:=\psi^*\gamma_0.$
The Lorentz condition has been treated in the C*--algebra context
(cf.~\cite{Lledo}) and it needs special treatment, e.g. indefinite metric
or nonregular states, but it is not very interesting for us, since it 
only affects the electromagnetic field ${\rm CCR}(S,\,B),$ hence
is independent of the charge $Q.$
The Maxwell equation is more interesting, since it involves both
factors of $\al B.$ and it expresses the interaction between the
two fields. It is however very difficult to enforce in the
C*--algebra context (and ultimately leads to the
conclusion that $\al B.$ is too
small an algebra to do this in). Naively, it seems that we can
easily realise both sides of the Maxwell equation in
the present C*--setting:
smear the left-hand side over $\al S.(\R^4,\,\R^4)$
\[
\int\Box A_\mu(x)\,f^\mu(x)\,dx=\int A_\mu\Box f^\mu dx
=A(\Box f)
\]
then this is realised in ${\rm CCR}(S,\,B)$ through the identification
of the generating Weyl unitaries $\delta_h$ with
the heuristic $\exp iA(h)$ where $h=\Box f.$
 If we smear the right-hand side 
 of the Maxwell equation:
\[
   j(f)=-e\int\wt\psi(x)\gamma_\mu\psi(x)f^\mu(x)\, dx\;,
\]
then $j(f)$ generates a Bogoliubov transformation $T_f$ on 
$L^2(\R^4,\,\C^4)$ by:
\begin{eqnarray*}
\Ad\big(\exp ij(f)\big)\psi(g)&:=&\left(\exp i\ad j(f)\right)\big(\psi(g)\big)\\[1mm]
&=&\psi(T_fg)=:\alpha\s T_f.(\psi(g))
\end{eqnarray*}
where $\alpha\s T_f.$ is its associated automorphism on 
${\rm CAR}(\al H.)$
(we will calculate $T_f$ explicitly in a simplified setting below).
Let $G\subset\aut\al B.$ be the discrete group generated in
$\aut\al B.$ by 
\[
\set\beta_f:=\alpha\s T_f.\otimes\iota, f\in{\al S.(\R^4,\,\R^4)}.
\]
and let $\nu$ denote its action on $\al B..$ Define the crossed product
\[
\al E.:=G\cross\nu.\al B.={\rm C}^*\set{\al B.,\,U_g},
U^*_g=U_g^{-1},\; \nu_g=\Ad U_g,\; U_gU_h=U_{gh},\;g,h\in G.
\]
then we identify the heuristic objects $\exp ij(f)$ with the implementing
unitaries $U\s\beta_f..$ So each side of the Maxwell equation has a C*--realisation,
and we only need to decide how to impose the constraint equation.
Heuristically, the Maxwell equations are imposed as state conditions:
$A(\Box f)\phi = j(f)\phi$ for vectors $\phi$ in the representing
Hilbert (or Krein) space. 
If we take instead the stronger condition 
$A(\Box f)^n\phi = j(f)^n\phi$ for $n\in\N,$ then we can rewrite
the constraint conditions in the form ${e^{iA(\Box f)}\phi}=
{e^{ij(f)}\phi}.$ This suggests that we choose constraint
unitaries $V_f:=U\s -\beta_f.\cdot\delta\s\Box f.$ in $\al E.$
and thus select our Dirac states $\omega$ on $\al E.$
by 
\[
\omega(V_f)=1\qquad\forall\; f\in {\al S.(\R^4,\,\R^4)}.
\]
As one expects from the interaction, this program encounters problems:
\begin{itemize}
\item[(1)] We always have that $\Box f\in\ker B,$ hence $\Box f$
corresponds to zero in $S$ (since we factor out by $\ker B).$
This can be remedied by changing $S$ to ${\al S.(\R^4,\,\R^4)},$
in which case $(S,\,B)$ is a degenerate symplectic space. 
This problem is connected to the fact that the heuristic smearing
formula
\[
A(f)=\int_{C_+}\left(a_\mu(\b p.)\wh{f}^\mu(p)
+a_\mu^+(\b p.)\bar{\wh{f}}^\mu(p)\right){d^3p\over p_0}
\]
cannot be correct for the interacting theory, since it implies that
$A(\Box f)=0\,,$ in contradiction with the Maxwell equation.
\item[(2)]
Interaction mixes the fermions and bosons, so it is unrealistic to expect that
the interacting fermion and boson fields will commute (as in the tensor 
product structure
of $\al B.).$ Even worse, perturbation theory suggests that 
the interacting fields need not be canonical, so the
assumption of the CCR and CAR relations 
for the interacting bosons and fermions is problematic.
\end{itemize}
\subsection{Model for the interacting Maxwell constraint}

Inspired by the observations above, we now propose an example
which is a simplified version of the Maxwell constraint.
Heuristically, we want to impose a constraint of the form
\[
a^*(x)\,a(x)=LA(x)
\]
where $a(x)$ is a fermion field on $\R^4,$ $A$ is a boson field
and $L$ is a linear differential operator on $\al S.(\R^4).$
To realise this, together with a superselection structure
 in a suitable C*--algebra setting, we present our construction
in six steps.\chop\medskip
STEP 1.\chop
\medskip\noindent
For the fermion field, let $\al H.=L^2(\R^4)$ and define ${\rm CAR}(
\al H.)$ in Araki's self-dual form (cf.~\cite{Ar}) as follows.
On $\al K.:=\al H.\oplus\al H.$ define an antiunitary involution
$\Gamma$ by $\Gamma(h_1\oplus h_2):=\ol h._2\oplus\ol h._1\;.$
Then ${\rm CAR}(\al H.)$ is the unique simple C*--algebra with
generators $\set\Phi(k),k\in{\al K.}.$ such that
$k\to\Phi(k)$ is antilinear, $\Phi(k)^*=\Phi(\Gamma k)\,,$
and
\[
\left\{\Phi(k_1),\,\Phi(k_2)^*\right\}=(k_1,\,k_2)\un\;,\qquad
k_i\in\al K.\;.
\]
The correspondence with the heuristic creators and annihilators
of fermions is given by $\Phi(h_1\oplus h_2)=a(h_1)+a^*(\ol h._2)\;,$
where
\[
a(h)=\int a(x)\,\ol h(x).\,d^4x\;,\qquad
a^*(h)=\int a^*(x)\, h(x)\, d^4x\;.
\]
\noindent STEP 2.\chop\medskip\noindent
For the boson field, let $S=\al S.(\R^4,\R),$
and let $K:S\to L^2(M,\mu)$ be a linear map, where $(M,\mu)$ is
a fixed measure space. Define a symplectic form on $S$ by
$B(f,g):={\rm Im}(Kf,\,Kg),$ where $(\cdot,\cdot)$ is the
inner product of $L^2(M,\mu).$ Note that $B$ is degenerate
if $\ker K$ is nonzero.
Define then ${\rm CCR}(S,\,B)=C^*\set\delta_f,f\in S.$
where the $\delta_f$ are unitaries satisfying the Weyl relations:
\[
\delta_f\cdot\delta_g=\delta\s f+g.\exp[iB(f,g)/2]
\]
i.e. ${\rm CCR}(S,\,B)$ is the $\sigma\hbox{--twisted}$
discrete  group algebra of $S$ w.r.t. the two--cocycle
$\sigma(f,g):=\exp[iB(f,g)/2]\,.$
\chop\medskip
STEP 3.\chop
\medskip\noindent
To combine the bosons and fermions in one C*--algebra, we want to
allow for the possibility that they may not commute with each other,
hence we will not take the tensor algebra 
${\rm CAR}(\al H.)\otimes{\rm CCR}(S,\,B)\,.$
However, we don't know what form their commutators should take, 
so we start with the free C*--algebra $\al E.$
generated by ${\rm CAR}(\al H.)$ and ${\rm CCR}(S,\,B)\,.$
The free C*--algebra $\al E.$ seems to be big enough
to allow for possible interactions, but it is also likely to contain
redundant elements. 

To be explicit, let $\al L.$ be the linear space spanned by
all monomials of the form $A_0B_0A_1B_1\cdots A_nB_n$ where
$A_i\in{\rm CAR}(\al H.)$ and $B_i\in{\rm CCR}(S,\,B)\,.$
Note that $\al L.$ is an algebra w.r.t. concatenation.
Factor out by the ideal generated by $\un\s{\rm CAR}.-
\un\s{\rm CCR}.$ and replace concatenation by multiplication
for any two elements in a monomial which are in the same
algebra (either ${\rm CAR}$ or ${\rm CCR}$) after the
factorisation.
Note that this will now produce all
possible monomials of elements in ${\rm CAR}(\al H.)$ and
${\rm CCR}(S,\,B)$ - just consider those monomials
in $\al L.$ with $A_0$ or $B_n$ the identity to obtain
all other monomials.
Now the resultant algebra $\al N.$ is a *--algebra with the involution
given by
\[
(A_0B_0\cdots A_nB_n)^*=B_n^*A_n^*\cdots B_0^*A_0^*\;.
\]
Form the enveloping C*--algebra $\al E.$ of $\al N.,$ i.e.
let
\[
\al I._0:=\bigcap\set\ker\pi,\pi\in
\hbox{Hilbert space representations of $\al N.$}.
\]
and set $\al E.:=\ol{\al N.\big/\al I._0}.$ where the closure
is w.r.t. the enveloping C*--norm, i.e.
\[
\|A\|:=\sup\set\|\pi(A)\|,\pi\in
\hbox{Hilbert space representations of $\al N.$}.\;.
\]
That $\al E.$ is nontrivial, follows from the fact that any
tensor product representation of ${\rm CAR}(\al H.)\otimes{\rm CCR}(S,\,B)$
defines a Hilbert space representation of $\al N.,$ hence it
follows that $\al E.$ is nonzero and that ${\rm CAR}(\al H.)$
and ${\rm CCR}(S,\,B)$ are faithfully embedded in $\al E.$
(as the images under the factorisation maps of the original
generating algebras in the construction).
Note that we have a surjective homomorphism $\zeta:\al E.\to
{{\rm CAR}(\al H.)\otimes{\rm CCR}(S,\,B)}$
given by
\[
\zeta(A_0B_0\cdots A_n B_n):= (A_0\cdots A_n)\otimes(B_0\cdots B_n)\;,
\quad A_i\in {\rm CAR}(\al H.),\; B_i\in{\rm CCR}(S,\,B)\,.
\]
Clearly the ideal $\al I._T$ of $\al E.$ generated by
the commutators $\big[{\rm CAR}(\al H.),\,{\rm CCR}(S,\,B)\big]$
is in $\ker\zeta.$ Since $\al E.$ probably contains redundant
elements, we do not require it to be simple.
$\zeta$ will be important in proofs below.
\chop\medskip
STEP 4.\chop
\medskip\noindent
Next, we would like to model in the curent C*--setting, the
global and local heuristic charges:
\[
Q=\int a^*(x)\,a(x)\,d^4x\,,\qquad
Q(f)=\int a^*(x)\,a(x)\,f(x)\,d^4x\,,\quad
f\in\al S.(\R^4,\R)\;.
\]
Let us calculate the Bogoliubov transformations which they induce:
\begin{eqnarray*}
&&\big[Q(f),\Phi(h_1\oplus h_2)\big] \\[1mm]
&=& \int\int\left[a^*(x)a(x)f(x),\,a(y)\ol h_1(y).+
a^*(y)\ol h_2(y).\right]\,d^4x\,d^4y \\[1mm]
&=& \int\int f(x)\bigg\{\Big(a^*(x)a(x)a(y)-a(y)a^*(x)a(x)\Big)
\ol h_1(y).\\[1mm]
&&\qquad+\Big(a^*(x)a(x)a^*(y)- a^*(y)a^*(x)a(x)\Big)\ol h_2(y).
\bigg\}d^4x\,d^4y \\[1mm]
&=& \int\int f(x)\bigg\{-\big\{a^*(x),\,a(y)\big\}a(x)\,\ol h_1(y).\\[1mm]
&&\quad \left(a^*(x)\big(\delta(x-y)-a^*(y)a(x)\big)
-a^*(y)a^*(x)a(x)\right)\ol h_2(y).\bigg\}\,d^4x\,d^4y\\[1mm]
&=&\int\int f(x)\bigg\{-\delta(x-y)\ol h_1(y).a(x) +
\delta(x-y)a^*(x)\ol h_2(y).\bigg\}\,d^4x\,d^4y\\[1mm]
&=& -a(\ol f.\cdot h_1)+a^*(f\cdot\ol h_2.)\\[1mm]
&=&\Phi(-\ol f.\cdot h_1\oplus\ol f.\cdot h_2)
=\Phi\big(f(-h_1\oplus h_2)\big)
\end{eqnarray*}
since $f$ is real. For the global charge $Q,$ just put 
$f=1$ in the last calculation. Thus
\begin{eqnarray*}
\big(\ad Q(f)\big)^n\big(\Phi(h_1\oplus h_2)\big)
&=&\Phi\left(f^n\cdot\big((-1)^nh_1\oplus h_2\big)\right)\qquad
\hbox{hence:}\\[1mm]
\left(\Ad\big(\exp\,iQ(f)\big)\right)\left(\Phi(h_1\oplus h_2)\right)
&=& \Big(\exp\,i\,\ad Q(f)\Big)\left(\Phi(h_1\oplus h_2)\right)\\[1mm]
&=& \sum_{n=0}^\infty{\big(i\,\ad\,Q(f)\big)^n\over n!}
\left(\Phi(h_1\oplus h_2)\right)\\[1mm]
&=& \sum_{n=0}^\infty{i^n\over n!}\,\Phi\Big(f^n\big((-1)^n
h_1\oplus h_2\big)\Big)\\[1mm]
&=& \Phi\big(e^{-if}h_1\oplus e^{if}h_2\big)=:\Phi\big(T_f(h_1\oplus h_2)
\big)\;.
\end{eqnarray*}
Now $T_f$ is unitary on $\al K.,$ and satisfies ${[T_f,\,\Gamma]}=0$
hence it is a Bogoliubov transformation (cf.~p43 in \cite{Ar}), and
so we can define automorphisms on 
${\rm CAR}(\al H.)$ by
\[
\wt\gamma_f\big(\Phi(k)\big):=\Phi(T_fk)\;.
\]
It is clear that $T_fT_g=T_{f+g}\,,$ hence that 
$\wt\gamma:\al S.(\R^4)+\R\to\aut({\rm CAR}(\al H.))$
is a homomorphism.
We extend these automorphisms to maps $\gamma_f$ on $\al E.$ by setting
\[
\gamma_f\restriction {\rm CAR}(\al H.) =\wt\gamma_f\;,\qquad
\hbox{and}\qquad
\gamma_f\restriction {\rm CCR}(S,B) = \iota
\]
where $\iota$ is the identity map.
The only relations between ${\rm CAR}(\al H.)$ and 
${\rm CCR}(S,B)$ in the free construction of $\al E.,$ is
$\un\s{\rm CAR}.=\un\s{\rm CCR}.,$ so since the definition of 
$\gamma_f$ preserves this relation, it will  extend to a well-defined
map on the free *--algebra $\al N..$ 
In fact, since $\gamma_f$ replaces ${\rm CAR}(\al H.)$
by an isomorphic one in a free construction, it will be 
an automorphism on $\al N.,$
and so will define an automorphism on the enveloping algebra $\al E..$

Let $G$ denote the Abelian group generated in
$\aut\al E.$ by $\set\gamma_f,f\in{\al S.}(\R^4)\cup\R.$
and equip it with the discrete topology.
Denote its action by $\beta:G\to\aut\al E.,$ and define the algebra
\[
\al A.:=G\cross\beta.\al E.\;,
\]
then we identify the implementing unitaries $U_{\gamma_f}\in\al A.$ of
$\gamma_f\in\aut\al E.$ with the heuristic objects
$\exp iQ(f),$ $f\in {\al S.}(\R^4)\cup\R$ (in the case that 
$f=t\in\R,$ we denote $Q(t)=tQ).$
Now $\gamma$ is a surjective homomorphism $\gamma:{\al S.(\R^4)+\R}
\to G$ and from the definitions above, it is clear that its kernel
is $2\pi\Z\subset\R,$ hence the discrete group $G$ is isomorphic to
$\al S.(\R^4)\times\T.$ Of course $\T$ will be our global gauge
group below.
\chop\medskip
STEP 5.\chop
\medskip\noindent
Next, we would like to realize in $\al E.$ the heuristic constraints
\[
Q(f)^n\psi=A(Lf)^n\psi\qquad\forall\;f\in{\al S.}(\R^4),\;\;
n\in\N
\]
where $L:S\to\ker K\subseteq\ker B$ is a given linear map.
First write the heuristic constraints in bounded form:
\[
e^{iQ(f)}\psi=e^{iA(Lf)}\psi\;,\quad\hbox{i.e.}\quad
e^{-iA(Lf)}e^{iQ(f)}\psi=\psi\;.
\]
So, given the identifications with heuristic objects above,
we define our constraint unitaries to be:
\[
\al U.:=\set\delta\s-Lf.\cdot U_{\gamma_f}=:V_f,f\in{\al S.}(\R^4,\R).
\subset\al A.\;.
\]
\begin{pro} 
\label{FCU}
$\al U.$ is first--class.
\end{pro}
The proof is in the next section.
The heuristic constraint conditions now correspond to the
application of the T--procedure to $\al U..$
\chop\medskip
STEP 6.\chop
\medskip\noindent
Now we will specify the superselection structure associated with
the global charge $Q$ using
the fact that $Q$ must take integer values on
the vacuum state. Recall that the global gauge transformations 
$\gamma\s t.\,,$ $t\in\R$ are implemented by the unitaries
$U\s\gamma_t.\in \al A.$ which we identify with the heuristic objects
$\exp itQ$ (cf. Step 3). For the superselection sectors
we need to find cyclic representations ${(\pi,\,\Omega)}$ 
such that 
\[
\pi(U\s\gamma_t.)\Omega=e^{itn}\Omega\qquad\forall\;t\in\R
\]
and some $n\in\Z$ (the heuristic corresponding conditions
are $Q\Omega=n\Omega).$
We recognise these as constraint conditions for Dirac states
of the constraint unitaries:
\[
\al V._n:=\set V_t^{(n)}:=e^{-itn}U\s\gamma_t.,t\in\R.\;.
\]
Denote the sets of these Dirac states by
\[
{\got S}_D^{(n)}:=\set\omega\in{\got S}(\al A.),
\omega(V_t^{(n)})=1\quad\forall\;t\in\R.\,.
\]
These folia of states will be our superselection sectors.
\begin{lem}
\label{SectDisj}
With notation as above, we have:
\begin{itemize}
\item[(i)] 
${\got S}_D^{(n)}\cap{\got S}_D^{(m)}=\emptyset$ if $n\not=m\;,$
\item[(ii)]
${\got S}_D^{(0)}\not=\emptyset\;.$
\end{itemize}
\end{lem}
\begin{beweis} (i)
If there is an $\omega\in{\got S}_D^{(n)}\cap{\got S}_D^{(m)}$
for $n\not=m,$ then
\begin{eqnarray*}
\omega\big(e^{-itn}U\s\gamma_t.\big)&=& 1\; =\;\omega\big(e^{-itm}U\s\gamma_t.\big) \\[1mm]
\hbox{so:}\qquad\omega\big(U\s\gamma_t.\big) &=& e^{itn}=e^{itm}\qquad\forall\;t
\end{eqnarray*}
which contradicts $n\not=m.$\chop 
(ii) In the proof of Lemma~\ref{FCU} we constructed a state 
$\omega_3\in{\got S}(\al A.)$ satisfying $\omega_3(U_g)=1$
for all $g\in G.$ If we take $g=\gamma\s t.,$ then this implies
that $\omega_3\in{\got S}_D^{(0)}.$
\end{beweis}
To connect with the usual machinery for superselection used above,
we need to exhibit the canonical endomorphisms (automorphisms in
the abelian case). We construct an action $\rho:\Z\to\aut\al A.$
such that its dual action on $\al A.^*$ satisfies
$\rho^*_k({\got S}_D^{(n)})={\got S}_D^{(n+k)}.$
\begin{defi} 
\label{RhDf}
For each $k\in\Z$ define a *-automorphism $\rho_k$ of $\al A.$ by:
\begin{eqnarray*}
\rho_k(A)=A\quad\forall\;A\in\al E.;\qquad
\rho_k(U\s\gamma_t.)=e^{itk}U\s\gamma_t.\quad\forall\,t\in\R;\\[1mm]
\rho_k(U\s\gamma_f.)=U\s\gamma_f.\quad\forall\;f\in\al S.(\R^4)\,.
\end{eqnarray*}
\end{defi}
\begin{lem}
\label{RkWD}
$\rho_k$ is well--defined, and $\rho_k\in\aut\al A.\,.$
\end{lem}
The proof is in the next section.
Recall that for any $\alpha\in\aut\al A.$ we define its dual
$\alpha^*:\al A.^*\to\al A.^*$ by $\alpha^*(f):=f\circ\alpha$
for all functionals $f\in\al A.^*.$
\begin{pro} 
With notation as above, we have $\rho^*_k({\got S}_D^{(n)})
={\got S}_D^{(n+k)}$ and ${\got S}_D^{(n)}\not=\emptyset$
for all $n\in\Z\;.$
\end{pro}
\begin{beweis}
Let $\omega\in\rho^*_k\big({\got S}_D^{(n)}\big),$
i.e. $\omega=\omega_n\circ\rho_k$ for some $\omega_n\in{\got S}_D^{(n)}.$
Thus
\[
\omega\left(e^{-it(n+k)}U_{\gamma\s t.}\right)
=\omega_n\left(e^{-it(n+k)}\rho_k\big(U_{\gamma\s t.}\big)\right)
=\omega_n\left(e^{-itn}U_{\gamma\s t.}\right) =1
\]
i.e. $\omega\in{\got S}_D^{(n+k)}\,.$
Conversely, for any $\omega\in{\got S}_D^{(n+k)}$ there is an
$\omega_n\in{\got S}_D^{(n)}$ for which
$\omega=\omega_n\circ\rho_k$ and it is obviously
$\omega_n=\omega\circ\rho_{-k}\,.$
Thus $\rho^*_k({\got S}_D^{(n)})
={\got S}_D^{(n+k)}.$
Since we have that ${\got S}_D^{(0)}\not=\emptyset,$
it is now immediate that ${\got S}_D^{(n)}=\rho^*_k({\got S}_D^{(0)})
\not=\emptyset\;.$
\end{beweis}
Recall from our earlier discussions that the canonical automorphisms
(Abelian case) must necessarily be outer on $\al A..$ 
\begin{pro}
\label{RhOut}
With notation as above, $\rho_k\in\out\al A.$ if $k\not=0\;.$
\end{pro}
The proof of this is long, and is in the next section.

From the action $\rho:\Z\to{\rm Out}\,\al A.$ we construct a
Hilbert extension (cf. Subsection~\ref{abelianHS}). First set
\[
\Lambda:=\set\Ad U\circ\rho_k,{U\in\al A.\;\hbox{unitary,}\;k\in\Z}.
\]
so $\Z\cong\Lambda\big/{\rm Inn}\,\al A.\;.$
So the class of $k\in \Z$ in $\Lambda\big/{\rm Inn}\,\al A.$
is $\chi\s k.:=\set\Ad U\circ\rho_k,{U\in\al A._u}.\,.$
Take the monomorphic section $\chi\s k.\to k,$
then it has a trivial cocycle $\sigma(n,\,m)=1$ for all $n,\,m\in\Z\,.$
Define $\al F.:=\Z\cross\rho.\al A.,$ then it has the dense *--algebra
\[
\al F._0:=\set\sum_{n\in F}A_nU^n,{A_n\in\al A.\,,\;
F\subset\Z\;\hbox{finite}}.
\]
where $U\in\al F.$ is the unitary 
which implements $\rho_1,$ i.e. $\rho_1=\Ad U\restriction\al A.\,.$
Fix $t\in\T=\wh\Z$ and define an action
$\alpha:\T\to\aut\al F.$ by
\[
\alpha_t\left(\sum_{n\in F}A_nU^n\right)
:=\sum_{n\in F}A_nt^nU^n\qquad\hbox{on $\al F._0\;.$}
\]
Then the fixed point algebra of $\alpha$ is $\al A.\,.$
We verify the compatibility
condition in Corollary~\ref{cor4.1}:
\begin{pro}
$\rho_k(\al D.)\sim\al D.$ for all $k\in\Z\,.$
\end{pro}
\begin{beweis}
The constraint unitaries from which we define $\al D.$ are
$V_f:=\delta\s-Lf.\cdot U_{\gamma_f}\,,$ $f\in\al S.(\R^4)\,.$
By definition~(\ref{RhDf}) we have $\rho_k\restriction\al E.
=\iota\,,$ hence $\rho_k(\delta_{-Lf})=\delta_{-Lf}\,.$
Also $\rho_k(U_{\gamma\s f.}) = U_{\gamma\s f.}$
for all $f\in\al S.(\R^4)\,,$ hence 
$\rho_k(V_f)=V_f$ for all $f\in\al S.(\R^4)\,.$
Thus $\rho_k$ preserves the Dirac states ${\got S}_D$
and hence $\rho_k(\al D.)=\al D.$ for all $k\in\Z\,.$
\end{beweis}
It remains to show that this Hilbert system is regular
and minimal.
However, at this stage we do not have a proof because little is known 
about the ideal $\al I._0$ factored out in Step~3.

\section{Proofs}
\label{Proofs}

\subsection*{Proof of Theorem~\ref{Teo.4.1}} 

{\bf (i)} 
We have that  
$\beta_g:=g\rest\of.$ 
The pointwise norm-continuity of
$\beta_{\al G.}$
follows from the pointwise norm-convergence topology of $\al G..$
So $\{\of,\al G.,\beta\}$ is a C*--dynamical system. Since
$\al A.$
is the fixed point algebra of 
$\al G.$,
the fixed point algebra of
$\beta\s{\al G.}.$ is $\of\cap\al A..$
By Theorem~\ref{Teo.2.12} we have that
$\of\cap\al A.= \al O..$\chop
If $A\in Z(\al A.),$ then ${[A,\al C.]}=0,$ hence $A\in\al O.$
by Theorem~\ref{Teo.2.2}(iii), and from this it follows that
$Z(\al A.)\subseteq Z(\al O.).$ \chop
{\bf (ii)} 
 Now let
$\al H._{\gamma}\subset\Pi_{\gamma}\al F.$.
If there is a unit vector $\Phi\in\al H._\gamma\cap\of,$
then by invariance of $\al H._\gamma\cap\of$ under $\al G.,$
we also have $\al H._\gamma\cap\of\supset\spa(\al G.\Phi)
=\al H._\gamma,$ where the last equality follows from
irreducibility of the action of $\al G.$ on $\al H._\gamma.$
Thus  $\al H._\gamma\cap\of \not=\{0\}$ implies that
$\al H._\gamma\subset\of,$  
hence
$\al H._\gamma\subset\Pi_\gamma\of.$

To prove that
$\Pi_{\gamma}\of=\hbox{clo-span}\,(\al O.\al H._{\gamma})$
we follow the proof of Lemma 10.1.3 in ~\cite{BW}. First, since
$\al O.=\Pi_\iota\of$,
it follows that
$\al O.\al H._\gamma\subseteq\Pi_\gamma\of$
by Remark~\ref{remark1}(v).

By Evans and Sund~\cite{ES},
$\Pi_\gamma\of$ is the closed span of all the $\al G.\hbox{--invariant}$
subspaces $\al E.\subset\of$ such that 
$\beta_\al G.$ 
acts on $\al E.$ as an element of $\gamma\in\wh{\al G./\al K.}.$
So for the reverse inclusion,
$\Pi_{\gamma}\of\subseteq\csp\{\al OH._\gamma\},$
it suffices
to show that $\spa\{\al OH._\gamma\}$ contains all
$\al G.\hbox{--invariant}$
subspaces 
$\al E.\subset\of$ 
such that $\al G.$
acts on $\al E.$ as an element of $\gamma.$
Let $\{\Psi_1,\ldots,\,\Psi_d\},$ $d={\rm dim}\,\gamma$
be a basis of such an $\al E.$ under which the matrix representation
of the action of $\al G.$ is an element of $\gamma,$ i.e.
\[
g\Psi_i=\sum_j\lambda_{ji}(g)\,\Psi_j
\]
where the matrix $(\lambda_{ji}(g))$ is a unitary matrix representation 
of $\al G.$ of the type $\gamma.$
Choose an orthonormal basis $\{\Phi_1,\ldots,\,\Phi_d\}$ of $\al H._\gamma$
which also transforms under $\al G.$ according to $(\lambda_{ji}(g)).$
Consider now the element $A:=\sum\limits_j\Psi_j\Phi^*_j\in\of.$ Then
\begin{eqnarray*}
g(A)&=&\sum_jg\left(\Psi_j\Phi^*_j\right)=
\sum_{i,\,k}\Big(\sum_j\lambda_{ij}(g)\,\ol\lambda_{kj}(g).\Big)
\Psi_i\Phi^*_k   \\[1mm]
&=& \sum_{i,\,k}\delta_{ik}\Psi_i\Phi^*_k =\sum_j\Psi_j\Phi^*_j=A \,.
\end{eqnarray*}
Thus $A\in\al O.,$ and hence all $\Psi_i=A\Phi_i\in\al OH._\gamma$
i.e. $\al E.\subset\spa(\al OH._\gamma).$
\chop
{\bf (iii)}
Let $\al H._\sigma$ have an orthonormal basis $\{\Phi_1,\ldots\Phi_d\}$
hence  $\rho_\sigma(F)=\sum\limits_{j=1}^d\Phi_jF\Phi_j^*$ for $F\in\al F.,$
$\rho_\sigma\rest\al A.=\sigma.$
Since $\{\Phi_j\}\subset\of=M(\df)$ it is clear that $\rho_\sigma$
preserves both $\df$ and $\of.$ Since $\rho_\sigma$ also preserves $\al A.,$
it preserves $\al D.=\df\cap\al A.$ and $\al O.=\of\cap\al A.,$
where these equalities come from Theorem~\ref{Teo.2.12}.

\subsection*{Proof of Theorem~\ref{Teo.4.2}}

\def\rh{\rho\s{\al H.}.}
{\bf (i)} 
Let $\al H.\subset\of$ have an orthonormal basis $\{\Phi_j\}.$
By the same proof as for Theorem~\ref{Teo.4.1}(iii) we have that
$\rh(\al D.)\subseteq\al D..$  \chop
Since $\of$ is a *--algebra and the relative multiplier algebra of
$\df\supset\al D.,$ we have that
\begin{eqnarray*}
[\Phi^*_j,\, D]&\in &\df\quad\hbox{for all}\;D\in\al D.,\;j. \\[1mm]
\hbox{Thus:}\qquad
\Phi_j[\Phi^*_j,\, D]&\in &\Phi_j\df=\rh(\df)\Phi_j
\subset\csp(\rh(\df)\al F.)  \\[1mm]
\hbox{i.e.}\qquad
D-\rh(D)&=&\sum_j\left(\Phi_j\Phi^*_jD-\Phi_jD\Phi_j^*\right)
\in\csp(\rh(\df)\al F.\,.)  \\[1mm]
\hbox{So}\qquad
D&\in &\csp(\rh(\df)\al F.)\quad\hbox{for all}\; D\in\al D.\,.
\end{eqnarray*}
Thus we have shown that $\al D.\subset\csp(\rh(\df)\al F.),$
and now we would like to show that 
${\csp(\rh(\df)\al F.)}={\csp(\rh(\al D.)\al F.)}.$
We have that 
\[
\csp(\al CF.)=\csp(\al DF.)=\csp(\df\al F.),
\]
so if we apply $\rh$ to both sides of the last equation, multiply by
$\al F.$ on the right and take closed span, we get:

\begin{eqnarray*}
\csp\big(\rh(\al D.)\rh(\al F.)\al F.\big) &=&
\csp\big(\rh(\df)\rh(\al F.)\al F.\big)  \\[1mm]
\hbox{i.e.}\qquad
{\csp(\rh(\df)\al F.)} &=&
\csp(\rh(\al D.)\al F.)\;.  \\[1mm]
\hbox{Thus:}\qquad\al D. \subset \csp(\rh(\al D.)\al F.) & &\quad
\hbox{and since $\al D.$ is a *--algebra in $\al A.,$}  \\[1mm]
\al D.\; \subseteq \;\csp(\rh(\al D.)\al F.)\!&\cap&\!
\csp(\al F.\rh(\al D.))\cap\al A.  \\[1mm]
&\subseteq &\csp(\al DF.)\cap\csp(\al FD.)\cap\al A.=\al D. 
\end{eqnarray*}
where we used $\df\cap\al A.=\al D..$ Thus
\[
\al D.=\csp(\rh(\al D.)\al F.)\cap\csp(\al F.\rh(\al D.))\cap\al A.
=\csp(\rh(\al D.)\al A.)\cap\csp(\al A.\rh(\al D.))
\]
which also follows from Theorem~\ref{Teo.2.12}, treating 
$\rh(\al D.)\subseteq\al D.$ as a second constraint set.
Thus $\al D.\sim\rh(\al D.)$ in $\al A..$

For the converse, let $\al D.\sim\rh(\al D.)$ and take 
$\Phi\in\al H..$ From the equation $\Phi D=\rh(D)\Phi$ for all
$D\in\al D.,$ we conclude that
\[
\Phi\cdot\csp(\al DF.) \subset \csp(\rh(\al D.)\al F.)=\csp(\al DF.)
\]
using $\al D.\sim\rh(\al D.).$
Since we have trivially that
$\Phi\cdot\csp(\al FD.)\subset\csp(\al FD.),$ it follows that
\[
\Phi\df=\Phi\left(\csp(\al FD.)\cap\csp(\al DF.)\right)
\subset\df
\]
so $\Phi$ is in the left multiplier of $\df.$
We also have that
\[
\csp(\al FD.)\Phi=\csp(\al F.\rh(\al D.))\Phi=\csp(\al F.\Phi\al D.)
\subseteq\csp(\al FD.)\;.
\]
Since trivially $\csp(\al DF.)\Phi\subseteq\csp(\al DF.),$ it
follows that
\[
\df\Phi=\left(\csp(\al FD.)\cap\csp(\al DF.)\right)\Phi
\subset\df
\]
and hence $\Phi$ is in the relative multiplier algebra 
of $\df,$ i.e. $\Phi\in\of$ by Theorem~\ref{Teo.2.2}(ii).\chop
{\bf (ii)}
Let $\al H._\sigma\subset\of\supset\al H._\tau,$
hence by (i) $\al D.\sim\sigma(\al D.)\sim\tau(\al D.).$

First let $X\in (\sigma,\,\tau)_{\al A.}\cap\al O.,$ i.e.
$X\in\al O.$ and $X\sigma(A)=\tau(A)X$ for all $A\in\al A..$
By letting $A$ range over only $\al O.\subset\al A.,$
we immediately get that $X\in{(\sigma\rest\al O.,\,\tau\rest\al O.)_{\al O.},}$
making use of Theorem 4.1(iii).
Therefore,
it suffices to prove that 
${(\sigma,\,\tau)_{\al A.}}\subset\al O..$

Let $X\in{(\sigma,\,\tau)_{\al A.}},$ i.e.
$X\in\al A.$ and
$X\sigma(A)=\tau(A)X$ for all $A\in\al A..$ Thus
\begin{eqnarray*}
X\cdot\csp(\al DA.)&=& X\cdot\csp(\sigma(\al D.)\al A.)
=\csp(X\sigma(\al D.)\al A.)   \\[1mm]
&\subseteq&
\csp(\tau(\al D.)X\al A.) \subseteq
\csp(\tau(\al D.)\al A.)=\csp(\al DA.)\;.
\end{eqnarray*}
Since we have trivially that 
$X\cdot\csp(\al AD.)\subseteq\csp(\al AD.),$
it follows that
\[
X\al D.\subseteq\csp(\al AD.)\cap\csp(\al DA.)=\al D.\,,
\]
i.e. $X$ is in the left multiplier of $\al D..$
Likewise:
\begin{eqnarray*}
\csp(\al AD.)\cdot X &=& \csp(\al A.\tau(\al D.))\cdot X
=\csp(\al A.\tau(\al D.)X)   \\[1mm]
&\subseteq&
\csp(\al A.X\sigma(\al D.))\subseteq\csp(\al A.\sigma(\al D.))  \\[1mm]
&=& \csp(\al AD.)\,.
\end{eqnarray*}
Since trivially $\csp(\al DA.)\cdot X\subseteq\csp(\al DA.),$
we have:
\[
\al D.\cdot X\subseteq\csp(\al AD.)\cap\csp(\al DA.)=\al D.\,,
\]
and hence $X$ is in the relative multiplier of $\al D.,$
i.e. $X\in\al O..$

\subsection*{Proof of Proposition~\ref{pro.4.7}}

(i) According to the decomposition
\[
\sigma(\cdot)=\sum_{j}V_{j}\rho_{\gamma_{j}}(\cdot)V_{j}^{\ast},\quad
V_{j}\in (\rho_{\gamma_{j}},\sigma)_{\al A.}
\]
we have
$\al H._{\sigma}=\sum_{j}V_{j}\al K.'_{j}$
where
$\rho_{\gamma_{j}}=\rho_{\al K.'_{j}}$
and 
the
$\al K._{j}'$
are irreducible w.r.t.
$\al G.$
carrying the representation
$\gamma_j\in\wh{\al G.}.$
Moreover
$\hbox{supp}\,\al K.'_{j}=1.$

Put
$E_{j}:=V_{j}V_{j}^{\ast}.$
Then
$\sum_{j}E_{j}=1.$
Since
$V_{j}\al K.'_{j}\subset\al H._{\sigma}\subset\of$
it follows that
\[
E_{j}=\hbox{supp}\,V_{j}\al K.'_{j}\in\al O.
\]
for all $j$. Therefore, by assumption, there are isometries
$W_{j}\in\al O.$
with
$E_{j}=W_{j}W_{j}^{\ast}.$
Now we put
\[
\al K._{j}:=W_{j}^{\ast}V_{j}\al K.'_{j}\subset\of.
\]
Then
$\al K._{j}$
is an algebraic Hilbert space with
$\hbox{supp}\,\al K._{j}=1$,
carrying the representation
$\gamma_{j}$
and we have
$V_{j}\al K.'_{j}=W_{j}\al K._{j}$.
Hence
$\al H._{\sigma}=\sum_{j}W_{j}\al K._{j}$
and
\[
\sigma(\cdot)=\sum_{j}W_{j}\rho_{\al K._{j}}(\cdot)W_{j}^{\ast},\quad
W_{j}\in (\rho_{\al K._{j}},\sigma)_{\al O.}
\]
follows. 

(ii) This follows from (i) using the  existence
of subobjects.

\subsection*{Proof of Theorem~\ref{Teo.4.3}}

Let
$\al H.\subset\al F.$
be an arbitrary algebraic Hilbert space. 
Then
$\xi(\al H.)\subset\al L.$
is also an algebraic Hilbert space.
with support 1. 
To see this, let 
$\{\Phi_{j}\}_{j}$
be an orthonormal basis of
$\al H.$,
i.e.
$\Phi_{j}^{\ast}\Phi_{k}=\delta_{j,k}1$
and
$\sum_{j}\Phi_{j}\Phi_{j}^{\ast}=1$
then the same relations are true for the system
$\{\xi(\Phi_{j})\}_{j}.$
In particular $\xi$ is 
injective on $\al H..$
Moreover, if
$\al H.$
is $\al G.$-invariant and
${g}(\Phi_{j})=\sum_{k}u_{k,j}(g)\Phi_{k}$
then
${g}^{\xi}\big(\xi(\Phi_{j})\big)=\sum_{k}u_{k,j}(g)\xi(\Phi_{k}),$
i.e.
$\xi(\al H.)$
carries the same representation as
$\al H..$
 In particular, if
$\al H._{\gamma}$
carries
$\gamma$,
i.e.
$\al H._{\gamma}\subset\Pi_{\gamma}\al F.$
then
$\xi(\al H._{\gamma})\subset\Pi_{\gamma}^{\xi}\al L.$.
This proves (ii) and (i).\chop
Let
$\al N._{\gamma}$ be an orthonormal basis for
$\xi(\al H._\gamma),$ then by the first part 
it is the image under $\xi$ of an
orthonormal basis $\{\Phi_{\gamma,j}\}_{j}$ of $\al H._\gamma.$
Let  $F=
\sum A_{\gamma,j}\Phi_{\gamma,j}\in\al F._{\rm fin}$
such that
$\xi(F)=0=\sum\xi(A_{\gamma,j})\xi(\Phi_{\gamma,j}).$
By applying $\al G.^\xi$ to this equality, and using
the relation 
$g^\xi(\xi(\Phi_{\gamma,j}))=\sum\limits_{k}u_{k,j}(g)\xi(\Phi_{\gamma,k})$
we get
$\sum\limits_{\gamma,j,k}u_{k,j}^{\gamma}(g)\xi(A_{\gamma,j})\xi
(\Phi_{\gamma,k})=0$ for all $g\in\al G..$  
Now the orthogonality relations for the matrix elements of the
irreducible representations of
$\al G.$
imply
$\xi(A_{\gamma,j})\xi(\Phi_{\gamma,k})=0$
for all
$\gamma\in\wh{\al G.},j,k$.
Hence
$\xi(A_{\gamma,j})=0$
follows. This proves (iii). From
$\xi\circ\Pi_{\gamma}=\Pi_{\gamma}^{\xi}\circ\xi$
(iv) follows.\chop
For (v) observe that the homomorphic images of isometries $V_i\in\al A.$
with $V_1V_1^*+V_2V_2^*=\un$ produces a pair of isometries in
$\xi(\al A.)$ satisfying the same relation. So Property B for $\al A.$
implies Property B for $\xi(\al A.)\,.$

\subsection*{Proof of Corollary~\ref{Cor.4.9}}

(i) Let $\lambda$ be generated by $\al H.$, i.e. let
$\lambda=\rho_{\al H.}$
such that
$\lambda(A)=\sum_{j}\Phi_{j}A\Phi_{j}^{\ast}$
where
$\{\Phi_{j}\}_{j}$
is an orthonormal basis of
$\al H.$.
Then
$\xi(\lambda(A))=\sum_{j}\xi(\Phi_{j})\xi(A)\xi(\Phi_{j})^{\ast}$
and
$\xi(A)=0$
implies
$\xi(\lambda(A))=0.$
Furthermore,
$\lambda^{\xi}(\xi(A))=\rho_{\xi(\al H.)}(\xi(A)).$\chop
(ii)$\lambda(\cdot)=\sum_{j}W_{j}\lambda_{j}(\cdot)W_{j}^{\ast}$
implies
$\lambda^{\xi}(\cdot)=\sum_{j}\xi(W_{j})\lambda_{j}^{\xi}(\cdot)\xi(W_{j})^{\ast}$
and
$(\lambda\circ\sigma)^{\xi}=\lambda^{\xi}\circ\sigma^{\xi}.$
Further, if
$\sigma(\cdot)=V^{\ast}\lambda(\cdot)V$
where
$V\in (\sigma,\lambda)$,
i.e.
$V\sigma(\cdot)=\lambda(\cdot)V$
then
$\xi(V)\sigma^{\xi}(\cdot)=\lambda^{\xi}(\cdot)\xi(V)$
and
$\sigma^{\xi}(\cdot)=\xi(V)^{\ast}\lambda^{\xi}(\cdot)\xi(V).$

In particular, if
$\lambda\cong\sigma$
then
$\lambda^{\xi}\cong\sigma^{\xi}.$

\subsection*{Proof of Theorem~\ref{Teo.4.4}}

(i) Let
$\sigma\in\hbox{Ob}\,\al T.^{\xi}$.
Then there is a $\al G.$-invariant algebraic Hilbert space
$\al H.\subset\al L.$
such that
$\sigma (X)=\sum_{j}\Psi_{j}X\Psi_{j}^{\ast},\;X\in\xi(\al A.),$
where
$\{\Psi_{j}\}_{j}$
denotes an orthonormal basis of
$\al H.$.
On the other hand, there is a corresponding $\al G.$-invariant
Hilbert space
$\al K.\subset\al F.$
such that
$\al H.$
and
$\al K.$
carry unitarily equivalent representations of
$\al G.$. 
In
$\al K.$
we choose an orthonormal basis
$\{\Phi_{j}\}_{j}$
such that the representation matrix of
$\al G.$
in
$\al H.$
w.r.t.
$\{\Psi_{j}\}_{j}$
coincides with that in
$\al K.$
w.r.t.
$\{\Phi_{j}\}_{j}.$
Then
$\xi(\al K.)$
transforms under
$\al G.$
w.r.t.
$\{\xi(\Phi_{j})\}_{j}$
with the same representation matrix. Now we put
\[
V:=\sum_{j}\Psi_{j}\xi(\Phi_{j})^{\ast}\in\al L..
\]
Obviously, $V$ is unitary and
${g}^{\xi}(V)=V$
for all
$g\in\al G.$,
i.e.
$V\in\xi(\al A.)$.
Then
$V\xi(\Phi_{j})=\Psi_{j}$
or
$\al H.=V\xi(\al K.)$
and
$\sigma=\hbox{Ad}\,V\circ\rho_{\al K.}^{\xi}.$

(ii) According to Corollary~\ref{Cor.4.9} and
$(\hbox{Ob}\,\al T.)^{\xi}\subseteq\hbox{Ob}\,\al T.^{\xi}$
the image
$\al C.^{\xi}$
of an equivalence class
$\al C.\subset\hbox{Ob}\,\al T.$
is contained in a unique equivalence class of
$\hbox{Ob}\,\al T.^{\xi}$.
But (i) says that every equivalence class
$\al E.$
of
$\hbox{Ob}\,\al T.^{\xi}$
is an image
$\al E.=\al C.^{\xi}$.

\subsection*{Proof of Lemma~\ref{arrow1}}

Let $A\in(\sigma,\,\tau)_{\al A.},$ then 
it follows immediately from
$A\sigma(B)=\tau(B)A,\;B\in\al A.$
that 
$\xi(A)\sigma^{\xi}(\xi(B))=\tau^{\xi}(\xi(B))\xi(A)$
for all $B\in\al A..$ Recall that $\xi(\al A.)$ 
is the fixed point algebra of $\al G.^\xi.$

\subsection*{Proof of Proposition~\ref{Prop.4.12}}

(i) This is obvious because the union
$\bigcup_{\gamma}\al N._{\gamma}$
of orthonormal bases
$\al N._{\gamma}$
of
$\al H._{\gamma}$
is an $\al A.$-left module basis of
$\al F._{\rm fin}$.

\noindent
(ii) 
By a straightforward calculation one obtains for all $F\in\al F.$ that:
\[
\langle\xi(F),\xi(F)\rangle_{\xi(\al A.)}=
\xi(\langle F,F\rangle_{\al A.})
\]
and
\[
\vert\xi(F)\vert_{\xi(\al A.)}=\Vert\xi(\langle F,F\rangle_{\al A.})
\Vert^{1/2}\leq\Vert\langle F,F\rangle_{\al A.}\Vert^{1/2}
=\vert F\vert_{\al A.},
\]
i.e.
$\xi$
is continuous w.r.t. the norm
$\vert\cdot\vert_{\al A.}.$
Now let
$F\in\clo\s{\vert\cdot\vert_{\al A.}}.(\ker\xi\cap
\al F._{\rm fin}),$ hence there is a sequence $\{F_n\}\subset
\ker\xi\cap\al F._{\rm fin}$ such that
$\vert F_{n}-F\vert_{\al A.}\to 0.$
Then
$\xi(F)=0$
follows. Conversely, let
$F\in\ker\xi$.
Recall
$\xi\circ\Pi_{\gamma}=\Pi_{\gamma}^{\xi}\circ\xi$
which implies
$\Pi_{\gamma}F\in\ker\xi$.
Now, according to Remark~\ref{remark1}~(iv) we have
$F=\sum_{\gamma}\Pi_{\gamma}F$
w.r.t. the
$\vert\cdot\vert_{\al A.}$-norm convergence. This implies
\[
F\in\hbox{clo}_{\vert\cdot\vert_{\al A.}}(\ker\xi\cap
\al F._{\rm fin}).
\]

\subsection*{Proof of Theorem~\ref{Teo.4.13}}

(i) Since $\HS$ is minimal and regular, there exists an
assignment 
$\sigma\to\al H._{\sigma}$
such that
an admissible (DR-)subcategory
$\al T._{\C}$
can be defined by
\[
(\sigma,\tau)_{\al A.,\C}:=(\al H._{\sigma},\al H._{\tau}),
\]
cf. Theorem~\ref{Teo1}. Now we use the morphism $\xi$ to define a
corresponding subcategory
$\al T.^{\xi}_{\C}$
for
$\xi(\al T.)$.
Recall
$\hbox{Ob}\,\xi(\al T.)=(\hbox{Ob}\,\al T.)^{\xi}\subset\hbox{Ob}\,
\al T.^{\xi}.$
We put
\[
\hbox{Ob}\,\al T.^{\xi}_{\C}:=\hbox{Ob}\,\xi(\al T.).
\]
Let
$\lambda,\sigma\in\hbox{Ob}\,\al T.$.
Then
$\lambda^{\xi},\sigma^{\xi}\in\hbox{Ob}\,\xi(\al T.)$
and the arrows are defined by
\[
(\sigma^{\xi},\tau^{\xi})_{\xi(\al A.),\C}:=
\xi((\sigma,\tau)_{\al A.,\C})=(\xi(\al H._{\sigma}),\xi(\al H._{\tau})).
\]
Then
\[
(\iota^{\xi},\iota^{\xi})_{\xi(\al A.),\C}
=\xi((\iota,\iota)_{\al A.,\C})=\xi(\C\un)=\C\xi(\un).
\]
It is straightforward to show that
$\al T.^{\xi}_{\C}$
has direct sums and subobjects (in the latter case note that if
$F$ is a nontrivial projection from
$(\sigma^{\xi},\sigma^{\xi})_{\xi(\al A.),\C}$
then there is a nontrivial projection 
$E\in (\sigma,\sigma)_{\al A.,\C}$
such that
$F=\xi(E)$
because the ($\al G.$-invariant) matrix
$\{p_{j,k}\}_{j,k}$
of $F$ w.r.t.
$\{\xi(\Phi_{\sigma,j})\}_{j}$,
where the
$\Phi_{\sigma,j}$
form an orthonormal basis of
$\al H._{\sigma}$,
can be used to define a corresponding $E$ in
$(\sigma,\sigma)_{\al A.,\C}$.
Furthermore, the permutation and conjugation structures of
$\al T._{\C}$
survive the morphism $\xi$.
Thus
$\al T.^{\xi}_{\C}$
is a DR-category. We use the notation
$\al T.^{\xi}_{\C}=\xi(\al T._{\C}).$
(This result means: The Hilbert system
$\{\xi(\al F.),\al G.\}$
is regular.)

(ii) First let
$\xi(\al A.)'\cap\xi(\al F.)=\xi(Z(\al A.)).$
Then, according to Theorem~\ref{Teo1}, property P.2 can be fulfilled by
an appropriate subcategory of the form described before. Second,
let property P.2 be satisfied. Then
$\xi(\al T._{\C})$
is an admissible (DR-)subcategory of
$\xi(\al T.)$.
Therefore, according to Theorem~\ref{Teo2} there is a corresponding
minimal and regular Hilbert extension
$\tilde{\al F.}$
of
$\xi(\al A.).$
The uniqueness part of Theorem~\ref{Teo2} gives that
$\tilde{\al F.}$
and
$\xi(\al F.)$
are $\al A.$-module isomorphic, hence
$\xi(\al A.)'\cap\xi(\al F.)=Z(\xi(\al A.))$
is also true. 

(iii) The inclusion
$\supseteq$
is obvious (see Lemma~\ref{arrow1}). The assertion is
\begin{equation}
\label{inclusion}
(\sigma^{\xi},\tau^{\xi})_{\xi(\al A.)}\subseteq\xi((\sigma,\tau)_{\al A.}).
\end{equation}
First we prove this inclusion for the admissible subcategory, i.e. we assert
\[
(\sigma^{\xi},\tau^{\xi})_{\xi(\al A.),\C}\subseteq 
\xi((\sigma,\tau)_{\al A.}).
\]
This is obvious by
\[
(\sigma^{\xi},\tau^{\xi})_{\xi(\al A.),\C}=
\xi((\sigma,\tau)_{\C})\subset\xi((\sigma,\tau)_{\al A.}).
\]
Second, recall that
$\xi((\sigma,\tau)_{\al A.
})$
is a right module w.r.t.
$\sigma^{\xi}(\xi(Z(\al A.)))$
and a left module w.r.t.
$\tau^{\xi}(\xi(Z(\al A.)))$.
On the other hand,
$(\sigma^{\xi},\tau^{\xi})_{\xi(\al A.)}$
is a right module w.r.t.
$\sigma^{\xi}(Z(\xi(\al A.)))$
and a left module w.r.t.
$\tau^{\xi}(Z(\xi(\al A.))).$
Further, according to P.2.
$(\sigma^{\xi},\tau^{\xi})_{\xi(\al A.),\C}$
is a generating subset for this module. Since, by assumption,
$Z(\xi(\al A.))$
and
$\xi(Z(\al A.))$
coincide, the inclusion~(\ref{inclusion}) follows.

(iv) This follows directly from
$\xi(\al A.)'\cap\xi(\al F.)=Z(\xi(\al A.))$
and the fact that the unitary equivalence classes of
$\al T.^{\xi}$
and
$\xi(\al T.)$
coincide.

\subsection*{Proof of Theorem~\ref{Teo.4.14}}

Since
$\al T._{\C}$
is an admissible (DR-)subcategory of
$\al T.$
we can apply Theorem\ref{Teo2}, i.e. there is a corresponding minimal
and regular Hilbert extension
$\HS$
of
$\al A.$.
Therefore the arrows of the category
$\al T._{\C}$
are given by
\begin{equation}
\label{arrowAdm}
(\sigma,\tau)_{\al A.,\C}=(\al H._{\sigma},\al H._{\tau}),
\end{equation}
where the Hilbert spaces
$\al H._{\sigma},\al H._{\tau}$
generate the endomorphisms
$\sigma,\tau$ respectively.

Now it is not hard to show that the morphism
$\xi$
can be extended to a morphism of
$\al F.$
by putting
\begin{equation}
\label{MorAdm}
\xi(\Phi_{\lambda,j}):=\Phi_{\lambda,j}
\end{equation}
where
$\lambda$
runs through a complete system of irreducible and mutually
disjoint endomorphisms and
$\{\Phi_{\lambda,j}\}$
denotes an orthonormal basis of the Hilbert space
$\al H._{\lambda}$
which generates
$\lambda$
(recall and use Proposition~\ref{Prop.4.12}). This morphism satisfies the
assumptions of Theorem~\ref{Teo.4.3}. The corresponding Hilbert system is
denoted by
$\{\al F.^{\xi},\al G.\}$
(recall that
$\al G.^{\xi}\cong\al G.$).
Equations
(\ref{arrowAdm}) and (\ref{MorAdm}) imply
\[
\sigma^{\xi}(\xi(\al A.))=\sum_{j}\Phi_{\lambda,j}\xi(A)\Phi_{\lambda,j}^{\ast}
\qquad\hbox{and}\qquad
\xi((\sigma,\tau)_{\al A.,\C})=(\sigma,\tau)_{\al A.,\C}.
\]
By assumption (iii) we have
$\xi((\sigma,\tau)_{\al A.}=(\sigma^{\xi},\tau^{\xi})_{\xi(\al A.)}$.
Since
$(\sigma,\tau)_{\al A.,\C}\subset(\sigma,\tau)_{\al A.}$
we have
$\xi((\sigma,\tau)_{\al A.,\C})\subset(\sigma^{\xi},\tau^{\xi})_
{\xi(\al A.)}.$
Therefore the subcategory
$\al T._{\C}^{\xi}$
of
$\xi(\al T.)$
defined by
\[
(\sigma^{\xi},\tau^{\xi})_{\xi(\al A.),\C}:=\xi((\sigma,\tau)_
{\al A.,\C}=(\sigma,\tau)_{\al A.,\C}
\]
is a DR-category.

Now we prove property P.2 for
$\al T._{\C}^{\xi}$. 
We have to show
\[
\sigma^{\xi}(Z(\xi(\al A.)))(\lambda^{\xi},\sigma^{\xi})_{\xi(\al A.),\C}
\lambda^{\xi}(Z(\xi(\al A.)))=(\lambda^{\xi},\sigma^{\xi})_{\xi(\al A.)}.
\]
The left hand side equals
\[
\sigma^{\xi}\xi(Z(\al A.))(\lambda^{\xi},\sigma^{\xi})_{\xi(\al A.),\C}
\lambda^{\xi}(\xi(Z(\al A.)))=
\xi(\sigma(Z(\al A.)))(\lambda^{\xi},\sigma^{\xi})_{\xi(\al A.),\C}
\xi(\lambda(Z(\al A.)))=
\]
\[
\xi(\sigma(Z(\al A.)))\xi((\lambda,\sigma)_{\al A.,\C})\xi(\lambda(Z(\al A.)))=
\xi(\sigma(Z(\al A.))(\lambda,\sigma)_{\al A.,\C}\lambda(Z(\al A.)))=
\xi((\lambda,\sigma)_{\al A.})
\]
and this coincides, by assumption, with the right hand side.

Now we can apply Theorem~\ref{Teo2} to obtain a further Hilbert extension
$\{\tilde{\al F.}^{\xi},\al G.^{\xi}\}$
where again
$\al G.^{\xi}\cong\al G.$.
Using the uniqueness part of Theorem~\ref{Teo2} we obtain that both
Hilbert extensions are
$\xi(\al A.)$-module isomorphic.

\subsection*{Proof of Proposition~\ref{FCU}} 

It suffices to show that there is one Dirac state, 
i.e. a state $\omega\in{\got S}(\al A.)$ with 
$\omega(\al U.)=1\;.$
Recall the homomorphism $\zeta:\al E.\to{\rm CAR}(\al H.)
\otimes{\rm CCR}(S,B)\;.$
Let $\omega_0\in{\got S}({\rm CAR}(\al H.))$ be that
quasi--free state which is zero on any normal--ordered
monomial of $a(f)$ and $a^*(h)$ of degree greater or equal
to 1. Then $\omega_0$ is invariant w.r.t. $\wt\gamma_f$
for all $f\in\al S.(\R^4)\cup\R.$
Moreover, since $L(S)\subset\ker B,$ there is a state
$\omega_1\in {\got S}({\rm CCR}(S,B))$ such that
$\omega_1(\delta\s Lf.)=1$ for all $f\in S.$
Then $\omega_2:=\omega_0\otimes\omega_1\in
{\got S}\big({\rm CAR}(\al H.)\otimes{\rm CCR}(S,B)\big)$
is a $\wt\gamma\s f.\otimes\iota\hbox{--invariant}$ state on 
${\rm CAR}(\al H.)\otimes{\rm CCR}(S,B)$ such that
$\omega_2(\un\otimes\delta\s Lf.)=1$ for all $f\in S.$
From this we define a state on $\al E.$ by 
$\wt\omega_2:=\omega_2\circ\zeta$ and since
$\zeta\circ\beta\s\gamma_f.=\wt\gamma\s f.\otimes\iota,$
it follows that $\wt\omega_2$ is 
$\beta\s G.\hbox{--invariant}$ on $\al E..$
Thus $\wt\omega_2$ extends to a state $\omega_3$
on $\al A.=G\cross\beta.\al E.$ by
$\omega_3(U_g)=1$ for all $g\in G,$
where $U_g$ denotes the unitary implementer for $\beta_g.$
So $\omega_3\in{\got S}(\al A.)$ is a Dirac state w.r.t.
the unitaries $U_G\cup\delta\s LS..$
Since the maximal set of constraint unitaries for 
a Dirac state is a group, it follows that for the products
$V_f=\delta\s-Lf.\cdot U_{\gamma_f}$ we have 
$\omega_3(V_f)=1$ for all $f,$ i.e. $\omega_3$
is a Dirac state w.r.t. $\al U.,$ hence $\al U.$ is first--class.

\subsection*{Proof of Lemma~\ref{RkWD}} 

Note that $\rho_k$ on the unitary implementers $\rho_k:U_G\to\al A.$
is a faithful group homomorphism. This is because it is the pointwise product of the
identity map $\iota$ with the character $\chi_k:U_G\to\C$ given by
$\chi\s k.\big(U\s\gamma_{f+t}.\big):=e^{itk},$ $t\in\R,$ $f\in\al S.(\R^4).$
Furthermore:
$\al A.=C^*\big(\rho_k(U_G)\cup\al E.\big).$
Thus the pair ${\{\rho_k(U_G),\;\al E.\}}$ is also a covariant system
for the action $\beta:G\to\aut\al E.$ (cf. Step~3), hence by the universal
property of cross--products (cf.~\cite{PR}) there is a *-homomorphism
$\theta:\al A.\to\al A.$ such that $\theta(A)=A$ for $A\in\al A.,$ and
$\theta(U_g)=\rho_k(U_g)\equiv\hbox{implementing unitary of the second system}.$
Then $\theta$ coincides with the definition of 
$\rho_k$ on the generating elements, so it follows that $\rho_k$
extends uniquely to a homomorphism. Since it is clear that $\rho_k$
is bijective (its inverse is $\rho_{-k})$ it follows that
$\rho_k$ is an automorphism of $\al A..$

\subsection*{Proof of Proposition~\ref{RhOut}} 

Proof by contradiction. Let $k\not=0$ and assume $\rho_k\in{\rm Inn}\,\al A.,$
i.e. $\rho_k=\Ad V$ for some unitary $V\in\al A.\,.$
Recall the homomorphism
\[
\zeta:\al E.\to
{\rm CAR}(\al H.)\otimes{\rm CCR}(S,\,B)
\]
encountered in Step 3. Since $(S,\,B)$ is degenerate, $\zeta(\al E.)$
is not simple which will be inconvenient in the proof below. Choose
therefore a maximal ideal $\al I.$ of ${\rm CCR}(S,\,B)$
(necessarily associated with a character of the centre
$Z\big({\rm CCR}(S,\,B)\big)),$ and let
\[
\eta:{\rm CAR}(\al H.)\otimes{\rm CCR}(S,\,B)\to
{\rm CAR}(\al H.)\otimes{\rm CCR}(S,\,B)\big/\al I.
\]
be the factorisation by the ideal $\un\otimes\al I.\,.$
Then the composition
\[
\xi:=\eta\circ\zeta :\al E.\to{\rm CAR}(\al H.)\otimes{\rm CCR}(S,\,B)\big/\al I.
\]
is a homomorphism of which the image is a simple algebra.\chop
Now the action $\beta:G\to\aut\al E.$ (Step 4) only affects
${\rm CAR}(\al H.)$ in $\al E.,$ so preserves the ideal generated
by the commutators ${[{\rm CAR}(\al H.),\,{\rm CCR}(S,\,B)]}$
in $\al E.$ as well as the ideal $\un\otimes\al I.\,.$
Thus each $\beta_g$ can be taken through the homomorphism $\xi$
to define an action
$\beta^\xi:G\to{\aut\big({\rm CAR}(\al H.)\otimes{\rm CCR}(S,\,B)\big/\al I.\big)}$
and it is just $\beta^\xi\s\gamma_f.=\wt\gamma\s f.\otimes \iota\,.$
Thus we can extend $\xi$ from $\al E.$ to $\al A.=
{G\cross\beta.\al E.}$ to get a surjective homomorphism
\[
\xi:\al A.\to G\cross\beta^\xi.\left({\rm CAR}(\al H.)\otimes{\rm CCR}(S,\,B)\big/\al I.\right)\;.
\]
Now $\rho_k\in\aut\al A.$ only affects $U_G,$ leaving $\al E.$ invariant, hence it
preserves $\ker\xi\subset\al A..$ Thus $\rho_k$ can be factored through $\xi$
to obtain the automorphisms $\rho^\xi_k\in\aut\xi(\al A.)$ by
\begin{equation}
\label{XVinn}
\rho^\xi_k\big(\xi(A)\big):= \xi\big(\rho_k(A)\big)\quad\forall\;A\in\al A.\,,\qquad
\hbox{and so}\qquad \rho^\xi_k=\Ad\xi(V)\;.
\end{equation}
Recall now that each element of the discrete crossed product
$\xi(\al A.)= {G\cross\beta^\xi.\xi(\al E.)}$
(with $\xi(\al E.)={\rm CAR}(\al H.)\otimes{\rm CCR}(S,\,B)\big/\al I.)$
can be written as a C*-norm convergent series $\sum\limits_{n=1}^\infty B_nU\s g_n.$
where $B_n\in\xi(\al E.)$ and $g_n\in G,$ (with $g_n$ distinct for different $n)$
and that the unitaries $U_G$
form a left $\xi(\al E.)\hbox{--module}$ basis.
In particular, for the implementing unitaries $\xi(V)$ of $\rho^\xi_k$ we have a
series $\xi(V)=\sum\limits_{n=1}^\infty B_nU\s g_n.\,,$ $B_n\in\xi(\al E.)\backslash 0.$
Since $\rho_k\restriction\al E.=\iota\,,$ it follows from equation~(\ref{XVinn})
that $\xi(V)A=A\xi(V)$ for all $A\in\xi(\al E.)\,,$ i.e.
\begin{eqnarray*}
A\xi(V)&=& \sum_{n=1}^\infty AB_nU\s g_n. \\[1mm]
&=&\xi(V)A=\sum_{n=1}^\infty B_nU\s g_n.A
=\sum_{n=1}^\infty B_n\beta^\xi\s g_n.(A)U\s g_n.
\end{eqnarray*}
for all $A\in\xi(\al E.)\,.$ So by the basis property of $U_G$ we have 
\begin{equation}
\label{BnIntertw}
AB_n=B_n\beta^\xi\s g_n.(A)\quad\forall\;A\in\xi(\al E.)=
{\rm CAR}(\al H.)\otimes{\rm CCR}(S,\,B)\big/\al I.\,.
\end{equation}
Since $\beta^\xi\s g_n.\restriction{\rm CCR}(S,\,B)\big/\al I.
=\iota,$ this implies that
$B_n\in\left({\rm CCR}(S,\,B)\big/\al I.\right)'.$ 
From the fact that ${\rm CCR}(S,\,B)\big/\al I.$
is simple (hence has trivial centre)
this means that $B_n\in{\rm CAR}(\al H.)\otimes\un,$
and hence equation~(\ref{BnIntertw}) claims that $B_n$
is a nonzero intertwiner between $\iota$ and
$\beta^\xi\s g_n.$ in ${\rm CAR}(\al H.).$
We next prove that $B_n$ is invertible, in which case
$\beta^\xi\s g_n.$ becomes inner on ${\rm CAR}(\al H.).$

Let $\pi:{\rm CAR}(\al H.)\to\al B.(\al L.)$ be any faithful irreducible
representation of ${\rm CAR}(\al H.)$ on a Hilbert space $\al L.$
e.g. the Fock representation), and let $\psi\in\ker\pi(B_n)\,.$
Then by (\ref{BnIntertw})
\[
\pi(B_nA)\psi=\pi(\beta\s g_n^{-1}.(A))\,\pi(B_n)\psi=0\qquad
\forall\;A\in{\rm CAR}(\al H.)\;.
\]
Thus $\pi({\rm CAR}(\al H.))\psi\subseteq\ker\pi(B_n)\,.$
However, in an irreducible representation every nonzero vector is cyclic,
so either $\psi=0$ or $\pi(B_n)=0,$ and the latter case is excluded 
by $B_n\not=0,$ $\pi$ faithful. Thus $\psi=0,$ i.e. we've shown that 
$\ker\pi(B_n)=\{0\}\,.$ Moreover by equation~(\ref{BnIntertw}) we have
\[
\pi(A)\pi(B_n)\varphi=\pi(B_n)\pi(\beta\s g_n.(A))\varphi
\quad\forall\;\varphi\in\al L.\backslash 0\,,\;A\in{\rm CAR}(\al H.)
\]
hence $\pi\big({\rm CAR}(\al H.)\big)\big(\pi(B_n)\varphi\big)
\subseteq{\rm Ran}\,\pi(B_n)$ for all $\varphi\in\al L.\backslash 0\,.$
Now $\pi(B_n)\varphi\not=0$ (by $\ker\pi(B_n)=\{0\})$ and so by Dixmier
2.8.4~\cite{Di} we have that
$\pi\big({\rm CAR}(\al H.)\big)\big(\pi(B_n)\varphi\big)=\al L.$
(no closure is necessary). Thus ${\rm Ran}\,\pi(B_n)=\al L.,$
i.e. $\pi(B_n)$ is invertible, and so since $\pi$ is faithful
(hence preserves the spectrum of an element) it follows that
$B_n$ is also invertible in ${\rm CAR}(\al H.).$

Using the fact that $B_n$ is invertible, equation~(\ref{BnIntertw})
becomes $\beta\s g_n.(A)=B_n^{-1}AB_n$ for all $A\in{\rm CAR}(\al H.)\,.$
Since $\beta\s g_n.$ is a *-homomorphism, this implies that
$B_n^{-1}A^*B_n=B_n^*A^*(B_n^{-1})^*,$ i.e.
$B_nB_n^*A^*=A^*B_nB_n^*$ for all $A\in{\rm CAR}(\al H.)\,,$
and since ${\rm CAR}(\al H.)$ has trivial centre, this means
$B_nB_n^*\in\C\un\,.$
Put $B_nB_n^*=:t_n$ (necessarily $t_n>0)$ then
$U_n:=B_n\big/\sqrt{t_n}$ satisfies
$U_nU_n^*=\un.$ By substituting $A$ by $\beta\s g_n^{-1}.(A)$
in (\ref{BnIntertw}) we also obtain $B_n^*B_n\in\C\un$ by the above
argument, then using $t_n=\|B_nB_n^*\|=\|B_n\|^2=\|B_n^*B_n\|=B_n^*B_n$
we get  also $U_n^*U_n=\un\,.$
Thus 
\[
\beta\s g_n.(A) =B_n^{-1}AB_n=\left({B_n\over\sqrt{t_n}}\right)^{-1}
A\left({B_n\over\sqrt{t_n}}\right)=U_n^*AU_n
\]
for $A\in {\rm CAR}(\al H.)\,,$ i.e. $\beta\s g_n.$ is inner on
${\rm CAR}(\al H.)\,.$
Recall however, that on ${\rm CAR}(\al H.)$ $\beta\s g_n.$ is just
an automorphism $\wt\gamma\s f_n.$ for some $f_n\in\al S.(\R^4)+\R,$
coming from a Bogoliubov transformation:
$\wt\gamma\s f_n.\big(\Phi(k)\big):=\Phi(T_{f_n}k)$ (cf. Step 4).
So for $\beta\s g_n.$ to be inner on ${\rm CAR}(\al H.),$
this means that either of $I\pm T_{f_n}$ must be trace--class
(cf. Theorem~4.1, p48 of Araki~\cite{Ar} or Theorem~4.1.4 in
\cite{PlR}).
However
\[
T_{f_n}(h_1\oplus h_2):=e^{-if_n}h_1\oplus e^{if_n}h_2\quad\forall\;
h_i\in\al H.=L^2(\R^4)\,.
\]
Now for any $f_n$ such that $T_{f_n}\not=I\,,$ it is clear
that the multiplication operators on $L^2(\R^4)$ by
$(I\pm e^{\pm if_n})$ cannot be trace--class.
This contradicts our finding that $\beta\s g_n.$ is inner 
if $g_n\not=e,$ hence only $g_n=e$ is possible in the series
$\xi(V)=\sum\limits_{n=1}^\infty B_nU\s g_n.$ i.e.
$\xi(V)= B\cdot U_e\,,$ $B\in\xi(\al E.)\backslash 0\,.$
But in this case equation~(\ref{BnIntertw}) becomes $AB=BA$
for all $A\in\xi(\al E.)$ and so since $\xi(\al E.)$ is simple,
$B\in\C\un\,.$ This however implies that 
$\iota=\Ad\xi(V)=\rho^\xi_k$ which cannot be because
$\rho_k(U\s\gamma_t.)=e^{ikt}U\s\gamma_t.$ factors unchanged through $\xi.$
From this contradiction, it follows that our initial assumption
$\rho_k\in {\rm Inn}\,\al A.$ is false.

\providecommand{\bysame}{\leavevmode\hbox to3em{\hrulefill}\thinspace}

\end{document}